\documentclass[11pt,preprint]{aastex}
\pdfoutput=1
\usepackage{graphicx}
\usepackage{amsmath}    % Advanced maths commands
\usepackage{amssymb}    % Extra maths symbols
\let\oldfootsep=\footnotesep
\newcommand\ltsima{$\; \buildrel <\over\sim \;$}
\newcommand\simlt{\lower.5ex\hbox{\ltsima}}
\newcommand\gtsima{$\; \buildrel >\over\sim \;$}
\newcommand\simgt{\lower.5ex\hbox{\gtsima}}

\newcommand\msun {M_\odot}

\newcommand\mearth {{M_\oplus}}

\shorttitle{}
\shortauthors{Bennett et al}

\begin{document}


\title{A Planetary Microlensing Event with an Unusually Red Source Star: MOA-2011-BLG-291}


\author{David P.~Bennett\altaffilmark{1,2,M},
Andrzej~Udalski$^{3,O}$,
Ian A.~Bond\altaffilmark{4,M},
Daisuke Suzuki$^{5,M}$,
Yoon-Hyun~Ryu$^{6,\mu}$,\\
and \\
Fumio Abe$^{7}$, 
Richard K.~Barry$^{1}$, 
Aparna Bhattacharya\altaffilmark{1,2},
Martin Donachie$^{8}$,
Akihiko Fukui$^{9,10}$, 
Yuki Hirao$^{1,2,11}$, 
%Y.~Itow$^{7}$,  
Kohei Kawasaki$^{11}$,
Iona Kondo$^{11}$,
Naoki Koshimoto$^{12,13}$,
Man Cheung Alex Li$^{8}$,
%C.H.~Ling$^{4}$, 
%K.~Masuda$^{7}$,  
Yutaka Matsubara$^{6}$, 
Shota Miyazaki$^{11}$,
Yasushi Muraki$^{6}$, 
Masayuki Nagakane$^{11}$,
Koji Ohnishi$^{14}$, 
%H.~Oyokawa$^{6}$, 
Cl\'ement Ranc$^{1}$,
Nicholas J.~Rattenbury$^{8}$, 
%Toshiharu~Saito$^{13}$,
Haruno Suematsu$^{11}$,
%Denis J.~Sullivan$^{15}$, 
Takahiro Sumi$^{11}$,
Paul~J.~Tristram$^{15}$,
Atsunori Yonehara$^{8}$,
%P.C.M.~Yock$^{7}$,
 \\ (The MOA Collaboration)\\
Micha{\l}~K.~Szyma{\'n}ski$^3$,
Igor~Soszy{\'n}ski$^3$,  
{\L}ukasz~Wyrzykowski$^3$,
Krzysztof~Ulaczyk$^{3,16}$,
Radek~Poleski$^{3,17}$,
Szymon~Koz{\l}owski$^3$,
Pawe{\l}~Pietrukowicz$^3$,
Jan~Skowron$^3$, \\ (The OGLE Collaboration)\\
Yossi~Shvartzvald$^{18}$,
Dan~Maoz$^{19}$,
Shai~Kaspi$^{19}$,
Matan~Friedmann$^{19}$, \\ (The Wise  Group)\\
Virginie~Batista$^{20}$,
Darren~DePoy$^{21}$,
Subo~Dong$^{22}$,
B.~Scott~Gaudi$^{17}$,
Andrew~Gould\altaffilmark{17,23,6},
Cheongho~Han$^{24}$
Richard~W.~Pogge$^{17}$,
Thiam-Guan~Tan$^{25}$,
Jennifer~C.~Yee$^{26}$, \\ (The $\mu$FUN Collaboration)
 } 
              

\affil{$^{1}$Code 667, NASA Goddard Space Flight Center, Greenbelt, MD 20771, USA;    \\ Email: {\tt david.bennett@nasa.gov}}
\affil{$^{2}$Department of Astronomy, University of Maryland, College Park, MD 20742, USA}
\affil{$^{3}$Warsaw University Observatory, Al.~Ujazdowskie~4, 00-478~Warszawa,Poland}
\affil{$^{4}$Institute of Natural and Mathematical Sciences, Massey University, Auckland 0745, New Zealand}
\affil{$^{5}$Institute of Space and Astronautical Science, Japan Aerospace Exploration Agency, Kanagawa 252-5210, Japan}
\affil{$^{6}$Korea Astronomy and Space Science Institute, Daejon 34055, Republic of Korea}
\affil{$^{7}$Institute for Space-Earth Environmental Research, Nagoya University, Nagoya 464-8601, Japan}
\affil{$^{8}$Department of Physics, University of Auckland, Private Bag 92019, Auckland, New Zealand}
\affil{$^{9}$Subaru Telescope Okayama Branch Office, National Astronomical Observatory of Japan, NINS,
3037-5 Honjo, Kamogata, Asakuchi, Okayama 719-0232, Japan}
\affil{$^{10}$Instituto de Astrof\'isica de Canarias, V\'ia L\'actea s/n, E-38205 La Laguna, Tenerife, Spain}
\affil{$^{11}$Department of Earth and Space Science, Graduate School of Science, Osaka University, Toyonaka, Osaka 560-0043, Japan}
\affil{$^{12}$Department of Astronomy, Graduate School of Science, The University of Tokyo, 7-3-1 Hongo, Bunkyo-ku, Tokyo 113-0033, Japan}
\affil{$^{13}$National Astronomical Observatory of Japan, 2-21-1 Osawa, Mitaka, Tokyo 181-8588, Japan}
\affil{$^{14}$Nagano National College of Technology, Nagano 381-8550, Japan}
%\affil{$^{13}$Tokyo Metropolitan College of Aeronautics, Tokyo 116-8523, Japan}
%\affil{$^{15}$School of Chemical and Physical Sciences, Victoria University, Wellington, New Zealand}
\affil{$^{15}$University of Canterbury Mt.\ John Observatory, P.O. Box 56, Lake Tekapo 8770, New Zealand}
\affil{$^{16}$Department of Physics, University of Warwick, Gibbet Hill Road, Coventry, CV4~7AL,~UK}
\affil{$^{17}$Dept.\ of Astronomy, Ohio State University, 140 West 18th Avenue, Columbus, OH 43210, USA}
\affil{$^{18}$IPAC, Mail Code 100-22, Caltech, 1200 E. California Blvd., Pasadena, CA 91125, USA}
\affil{$^{19}$School of Physics and Astronomy, Tel-Aviv University, Tel-Aviv 69978, Israel}
\affil{$^{20}$Institut d'Astrophysique de Paris, 98 bis bd Arago, 75014 Paris, France}
\affil{$^{21}$Department of Physics, Texas A\&M University, 4242 TAMU, College Station, TX 77843-4242, USA}
\affil{$^{22}$Kavli Institute for Astronomy and Astrophysics, Peking University, Beijing 100871, China}
\affil{$^{23}$Max-Planck-Institute for Astronomy, K\"onigstuhl 17, 69117 Heidelberg, Germany}
\affil{$^{24}$Department of Physics, Chungbuk National University, Cheongju 361-763, Republic of Korea}
\affil{$^{25}$Perth Exoplanet Survey Telescope, Perth, Australia}
\affil{$^{26}$Harvard-Smithsonian Center for Astrophysics, 60 Garden Street, Cambridge, MA 02138 USA}
\affil{$^{M}$MOA Collaboration}
\affil{$^{O}$OGLE Collaboration}
\affil{$^{\mu}$MicroFUN Collaboration}

%\clearpage


\begin{abstract}
We present the analysis of planetary microlensing event MOA-2011-BLG-291, which
has a mass ratio of $q=(3.8\pm0.7)\times10^{-4}$ and
a source star that is redder (or brighter) than the bulge main
sequence. This event is located at a low Galactic latitude in the survey area that is
currently planned for NASA's WFIRST exoplanet microlensing survey. This unusual color for a
microlensed source star implies that we cannot assume that the source star is in the
Galactic bulge. The favored interpretation is that the source star is a lower main sequence star
at a distance of $D_S=4.9\pm1.3\,$kpc in the Galactic disk.
However, the source could also be a turn-off star on the far side
of the bulge or a sub-giant in the far side of the Galactic disk if it experiences
significantly more reddening than the bulge red clump stars. However, these possibilities
have only a small effect on our mass estimates for the host star and planet.
We find host star and planet masses of $M_{\rm host} =0.15^{+0.27}_{-0.10}\msun$
and $m_p=18^{+34}_{-12}\mearth$ 
from a Bayesian analysis with a standard Galactic model under the assumption 
that the planet hosting probability does not depend on the host mass or distance.
However, if we attempt to measure the host and planet masses with host star
brightness measurements from high angular resolution follow-up imaging, the 
implied masses will be sensitive to the host star distance. The WFIRST exoplanet
microlensing survey is expected to use this method to determine the masses
for many of the planetary systems that it discovers, so this issue has important design
implications for the WFIRST exoplanet microlensing survey.
\end{abstract}

%\clearpage

\section{Introduction}
\label{sec-intro}
The exoplanet microlensing survey \citep{bennett18_wfirst} of
NASA's Wide Field Infrared Survey Telescope (WFIRST) \citep{WFIRST_AFTA} offers several 
substantial advantages over ground-based microlensing surveys for the study of 
extrasolar planetary systems. The primary advantages are due to the higher angular
resolution, which allows the detection of sub-Earth-mass planets 
over a wide range of separations \citep{bennett96,bennett02,penny18}. WFIRST's high angular resolution 
also enables the direct detection of the planetary host stars, which can be used to determine their masses 
\citep{bennett06,bennett07,bennett15,bennett16,batista14,batista15,dong-ogle71,fukui15,kosh17_ob120950}. 
This is important because the masses are often not available for exoplanetary systems
discovered by microlensing. WFIRST's wide field infrared focal plane also provides a significant
advantage \citep{bennett_MPF} over the optical focal planes that are currently used by 
ground-based surveys \citep{sako_moacam3,ogle4,kmtnet}. The source stars
in the Galactic bulge, which offer the highest observable microlensing rate of
any area of the sky, provide an even higher microlensing rate in the infrared,
due to the high dust extinction in the foreground of the bulge.

The event, MOA-2011-BLG-291, that we analyze in this paper, at Galactic coordinates of
$(l, b) = 0.9015^\circ, -1.9693^\circ$, is located in or near the candidate fields
for the WFIRST microlensing survey. So, the unusually red source that we determine
for this event's source star is something that might be common for planetary events
discovered by WFIRST. In fact, there already seems to be evidence for this. Of
the 60 published planetary microlensing events, there are 16 located at a Galactic latitudes of 
$|b| < 2.1$. For three of these events \citep{bennett12,mroz17_2pl,shvartzvald18}, 
there is no color measurement,
but 4 of the remaining 13 low latitude events have anomalously red sources. Besides
MOA-2011-BLG-291, these are OGLE-2013-BLG-0341 \citep{gould14}, 
OGLE-2013-BLG-1761 \citep{hirao17}, and OGLE-2014-BLG-0676 \citep{rattenbury17}.
Three more have colors that are marginally redder than the main sequence or subgiant branch
\citep{mroz17_ob160596,hwang18,ranc18}.
While the color measurements for these events are sometimes challenging due to
high extinction, it is unlikely to be a coincidence that 30\% of these low latitude events
are redder than the bulge main sequence or the bulge subgiant branch in the case of
OGLE-2013-BLG-0341.

There are two obvious ways in which we might expect that low latitude events would be
more likely to have anomalously red sources. The low latitude lines-of-sight stay much
closer to the Galactic plane than higher latitude directions, and so they encounter a
higher density of foreground Galactic disk stars. They are brighter than bulge stars
of the same spectral type because they are closer, while they are likely to be behind most of
the dust in the Galactic disk. These stars are not expected to experience significantly
less extinction than the bulge stars, because the scale heigh of dust in the Galactic
disk is much smaller than the stellar scale height \citep{drimmel}. As a result,
they appear above the bulge stars on the color magnitude diagram (CMD),
but this means that they also appear redder, because intrinsically faint main sequence
stars are redder than brighter main sequence stars.

The dust scale height is known to be low in the stellar neighborhood, and for most
lines of sight to the Galactic bulge, we observe a tight red clump feature in the 
CMD.  This suggests that there is little extinction in the bulge itself, and this conclusion
is bolstered by observations of external galaxies, which usually appear to have
little dust in their central bulges. However, we have little direct evidence regarding 
the possibility of dust beyond $\sim 9\,$kpc on low latitude lines-of-sight through
the bulge. So, this is a possibly that we consider in this paper.

This paper is organized as follows. In Section~\ref{sec-lc_data} we describe the
light curve data and photometry. We also discuss the real time modeling effort that
failed to find a convincing planetary signal and the retrospective analysis that
confirmed that this was a planetary microlensing event.
In Section~\ref{sec-lc}, we describe the light curve modeling and present the
best fit models. We describe the photometric calibration of the OGLE and MOA
data and the determination in Section~\ref{sec-radius}. We then 
derive the lens system properties in Section~\ref{sec-lens_prop} including
some speculative possibilities involving excess extinction beyond $9\,$kpc. In
Section~\ref{sec-conclude}, we discuss the implications of our analysis for the
WFIRST mission and reach our conclusions.

\section{Light Curve Data and Photometry}
\label{sec-lc_data}

Microlensing event MOA-2011-BLG-291, at ${\rm RA} =17$:55:28.29, 
${\rm  DEC} = -29$:10:14.4, and Galactic coordinates $(l, b) = (0.9015, -1.9693)$, was 
identified and announced as a microlensing candidate by the Microlensing Observations
in Astrophysics (MOA) Collaboration Alert system \citep{bond01} 
on 3 July 2011. The Microlensing Follow-up Network ($\mu$FUN) issued a high
magnification alert two days later, but the follow-up groups were unable to 
obtain much photometry at the peak. Fortunately, this event was in the area
of the sky monitored by three different survey teams. In addition to MOA, it 
was also observed by the Optical Gravitational Lensing Experiment (OGLE) Collaboration
as a part of the OGLE-IV survey \citep{ogle4} and the
Wise microlensing survey \citep{shvartzvald16}. The  $\mu$FUN group did obtain
data from the Perth Exoplanet Survey Telescope (PEST),
and the 1.3m SMARTS telescope at the Cerro Tololo Interamerican Observatory
(CTIO). 

Photometry of the MOA data was performed with the MOA pipeline \citep{bond01}, which
also employs the difference imaging method \citep{tom96}. The OGLE Collaboration provided 
optimal centroid photometry using the 
OGLE difference imaging pipeline\citep{ogle-pipeline}. The Wise data were reduced using the 
Pysis difference
imaging code \citep{albrow09}, and the $\mu$FUN CTIO and PEST data were reduced with
DoPHOT \citep{dophot}.

There were several reports of possible light curve anomalies at the time of the event, but
there were no light curve models that were widely circulated immediately 
after the event. However, the planetary nature of the event was established
during the 2013 re-analysis of a MOA microlensing events that led to the MOA-II statistical
analysis of exoplanets found by microlensing \citep{suzuki16}. This re-analysis also
led to the discovery of 3 other planetary microlensing events, MOA-2008-BLG-379
\citep{suzuki14,suzuki14e}, OGLE-2008-BLG-355 \citep{koshimoto14}, 
and MOA-2010-BLG-353 \citep{rattenbury15}.

\begin{figure}
\epsscale{0.9}
\plotone{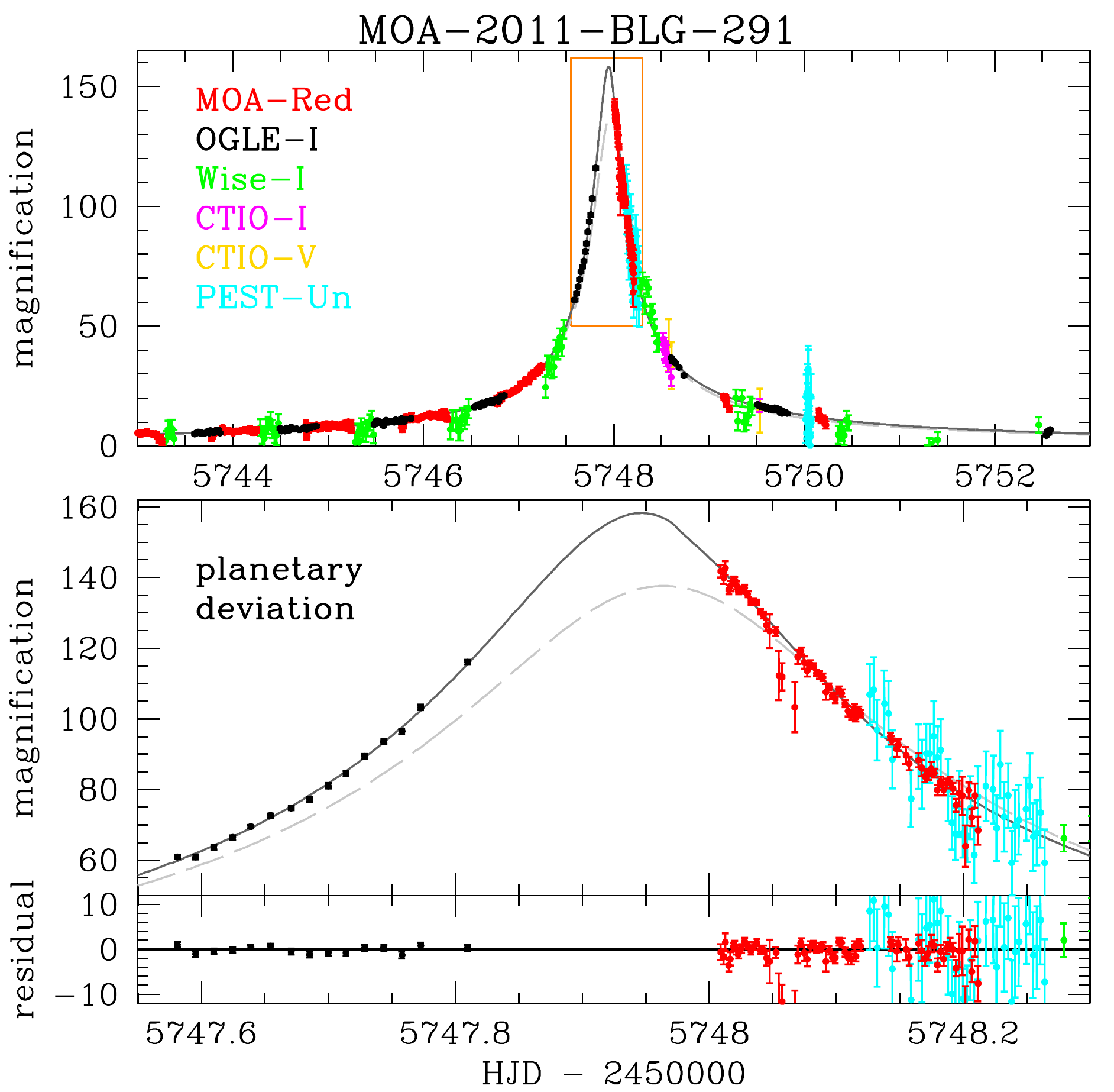}
\caption{The best binary lens model for  the MOA-2011-BLG-291 light curve.
The MOA-red data are shown in red while the OGLE, Wise, and CTIO 
$I$-band data are shown in black, green and magenta, respectively. The solid line is the best fit
model, while the grey dashed line is the single lens model with the same parameters as the
best fit model.
\label{fig-lc}}
\end{figure}

\section{Light Curve Models}
\label{sec-lc}

Our light curve modeling was done using the image centered ray-shooting method
\citep{bennett96} with the initial condition grid search method described in
\citet{bennett-himag}. The best fit planetary light curve model is shown in Figure~\ref{fig-lc},
with parameters given in Table~\ref{tab-mparams}. The parameters that this model
has in common with a single lens model are the Einstein radius crossing time, $t_E$, 
and the time, $t_0$, and distance, $u_0$, of closest approach between the lens center-of-mass 
and the source star. For a binary or planetary lens system, there is also the mass ratio 
of the secondary to the primary lens,
$q$, the angle between the lens axis and the source trajectory, $\alpha$, and the
separation between the lens masses, $s$. 
\begin{deluxetable}{ccccc}
\tablecaption{Model Parameters
                         \label{tab-mparams} }
\tablewidth{0pt}
\tablehead{
%% Use a footnote to explain numbering.
%& & & & \multicolumn{2}{c} {MCMC averages} \\
\colhead{parameter}  & \colhead{units} &
\colhead{$s\sim 1.2$} & \colhead{$s\sim 1.1$} &\colhead{MCMC averages}
}  % end header.

\startdata

$t_E$ & days & 23.645 & 22.958 & $23.5\pm 0.7$  \\
$t_0$ & ${\rm HJD}-2455700$ & 47.9641 & 47.9539 &  $47.963\pm 0.003$  \\
$u_0$ & & -0.007265 & -0.007237 & $-0.00729\pm 0.00027$  \\
$s$ & & 1.20828 & 1.10671 & $1.197 \pm 0.025$  \\
$\alpha$ & radians & 3.07475 & 3.04013 & $3.072\pm 0.010$  \\
$q$ & $10^{-4}$ & 4.0933 & 1.4017 & $3.80 \pm 0.70$  \\
$t_\ast$ & days & 0.15115 & 0.13904 & $0.148\pm 0.007$ \\
$I_s$ & & 20.742 & 20.716 & $20.747\pm 0.035$  \\
$V_s$ & & 23.475 & 23.500 & $23.468\pm 0.071$  \\
fit $\chi^2$ &  & 17545.08 & 17557.60 &  \\

\enddata
\end{deluxetable}

The length parameters, $u_0$ and $s$, are normalized by the Einstein radius of this total system mass, 
$R_E = \sqrt{(4GM/c^2)D_Sx(1-x)}$, where $x = D_L/D_S$ and $D_L$ and $D_S$ are
the lens and source distances, respectively. ($G$ and $c$ are the Gravitational constant
and speed of light, as usual.) 
For every passband, there are two parameters to describe the unlensed source
brightness and the combined brightness of any unlensed ``blend" stars that are
superimposed on the source. Such ``'blend" stars are quite common because
microlensing is only seen if the lens-source alignment is $\simlt \theta_E \sim 1\,$mas,
while stars are unresolved in ground based images if their separation is
$\simlt 1^{\prime\prime}$. 
% However, with ground-based seeing, the background contains '
% many unresolved stars, so it is uneven. This makes it possible for realistic cases of ``negative
% blending\rlap." More artificial negative blending can occur with difference imaging photometry
% that does not attempt to identify a source star in the reference image. In any case, these 
These
source and blend fluxes are treated differently from the other parameter because
the observed brightness has a linear dependence on them, so for each set of 
nonlinear parameters, we can find the source and blend fluxes that minimize the
$\chi^2$ exactly, using standard linear algebra methods \citep{rhie_98smc1}.

\begin{figure}
%\epsscale{0.7}
\plotone{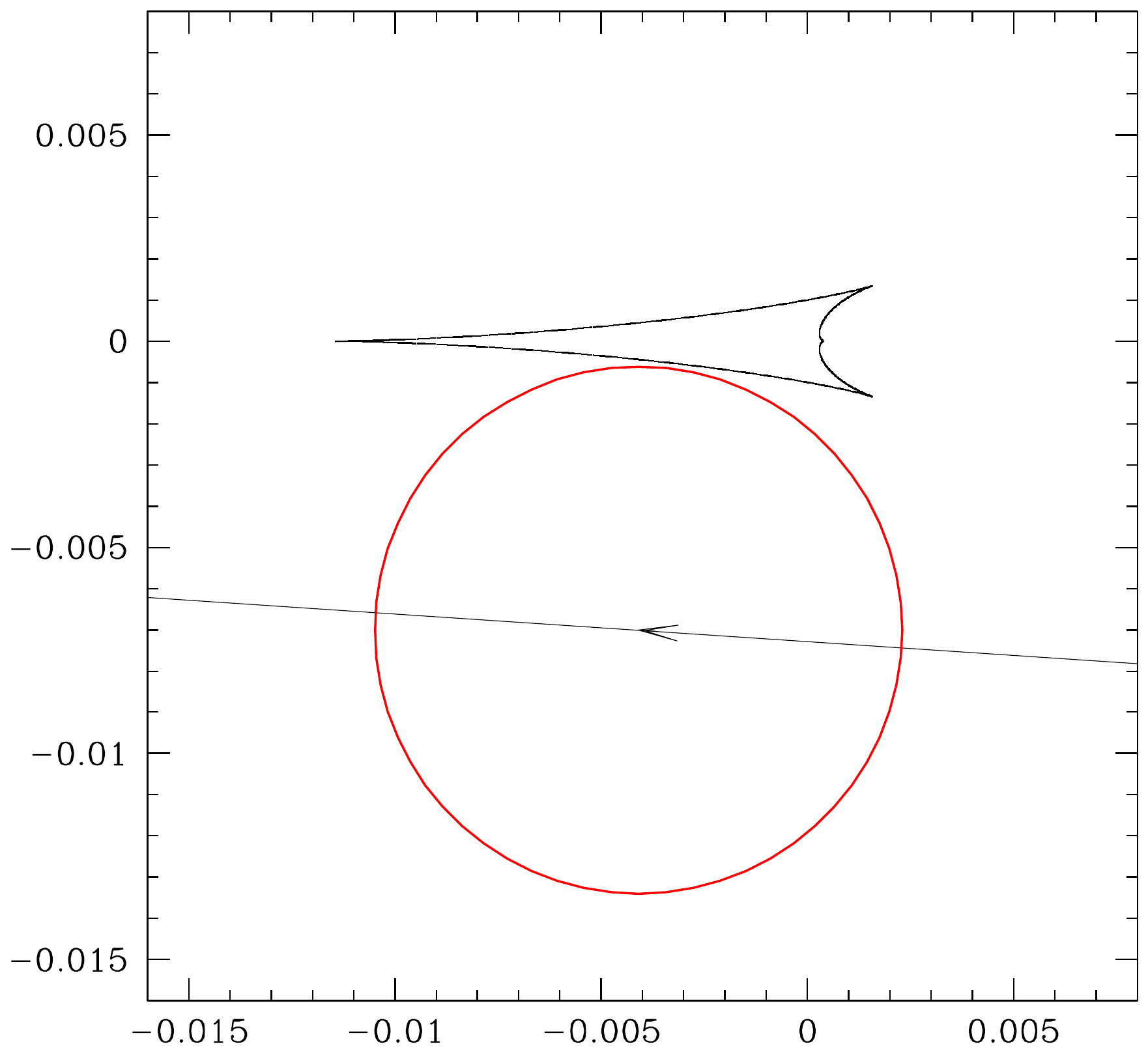}
\caption{The caustic configuration for the best fit model is plotted in units of the 
Einstein radius. The line with the arrow represents the motion of the center of the source
star, and the red circle indicates the size of the source star.
\label{fig-caustic}}
\end{figure}

The best fit model gives a $\chi^2$ improvement over the best single lens model of
$\Delta\chi^2 = 369.23$, and this $\chi^2$ difference is almost entirely in the OGLE and MOA data,
which dominate the coverage of the light curve peak.

Figure~\ref{fig-caustic} shows the caustic and source trajectory for the
best fit model. This (and the $\alpha \approx \pi$ parameters values given in Table~\ref{tab-mparams})
indicates that the source trajectory is nearly parallel to the lens axis. 
Table~\ref{tab-mparams} also gives the parameters of a second model that is worse than
the best fit model by $\Delta\chi^2 = 12.52$. This model is very similar to the best fit model,
a source trajectory that nearly grazes a central caustic of about the same size as the central
caustic of the best fit model, shown in Figure~\ref{fig-caustic}. However, in this case, the
mass ratio is about three times smaller, at $q = 1.4017\times 10^{-4}$, than the best fit value
of $q = 4.0933\times 10^{-4}$.  and the separation is closer to $s = 1$.

This means that
the source trajectory is likely to pass close to the planetary caustics. So, we might
possibly have a second signal from the same planet, although the signal would be at
much lower magnification. Events with strong signals from both the central and
planetary caustics can also give strong microlensing parallax signals even though
they may be of relatively short duration \citep{sumi16}. Unfortunately, the source star
for MOA-2011-BLG-291 is quite faint, and there is no significant detection of a planetary
caustic signal at lower magnification. So, we do not detect a significant microlensing parallax signal.

\section{Photometric Calibration and Source Radius}
\label{sec-radius}

\begin{figure}
\epsscale{0.9}
%\plotone{cmd_mb11291_VIhstrS3.pdf}
\plotone{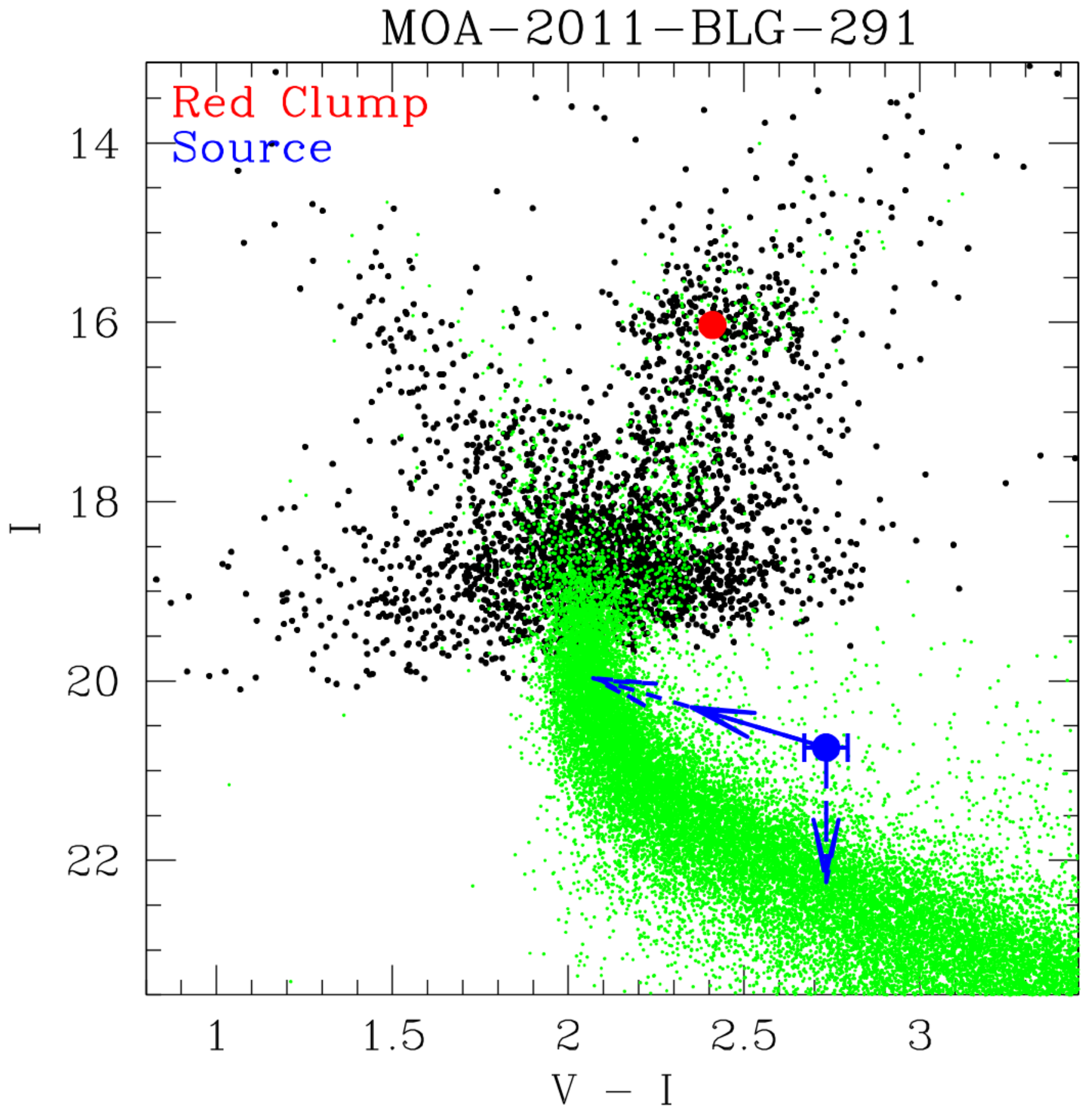}
\caption{The $(V-I,I)$ color magnitude diagram (CMD) of the OGLE-IV stars within 
$120^{\prime\prime}$ of MOA-2011-BLG-291 transformed to calibrated
Johnson $V$ and Cousins $I$ using the transformation given by \citet{ogle4} with the zero points
reported in the text. The red spot indicates
red clump giant centroid, and the blue indicates the source magnitude
and color. The green dots represent the HST Baade's Window CMD of \citet{holtzman98} transformed
to the extinction and Galactic longitude appropriate for this field. The blue arrows indicate ways
to put the source in more densely populated regions of the CMD. If the source has roughly the
same extinction as the red clump, then it could be a foreground mid-K dwarf at $D_S \sim 4\,$kpc, so 
we would move it downward by about 1.4 mag to reach the bulge. Alternatively, if it suffers more extinction
than the bulge stars, it could be a sub-giant or main sequence star beyond the bulge.
\label{fig-cmd}}
\end{figure}

Because light curve models listed in Table~\ref{tab-mparams} constrain the finite source
size through measurement of the source radius crossing time, $t_*$,  we can derive the angular Einstein
radius, $\theta_E = \theta_* t_E/t_*$, if we know the angular size of the source star, $\theta_*$.
This can be derived from the extinction corrected brightness and color of source star
\citep{kervella_dwarf,boyajian14}. Unfortunately, we do not have $V$-band measurements at a
high enough magnification to give us a reliable color measurement, so we must use the 
difference between the OGLE-$I$ and MOA-red passbands to determine the color. This
target is not in the OGLE-III survey footprint, so we calibrate to OGLE-IV photometry.
While an OGLE-IV photometry catalog has not been published, the color terms are given
in Table~1 of \citet{ogle4}. The zero points for OGLE-IV field BLG505.24 are 
$\Delta{\rm ZP}_I = -0.01$ and $\Delta{\rm ZP}_V = 0.19$, which can be inserted into 
equation~1 of \citet{ogle4} to derived calibrated magnitudes. Combining this relation with
the relation that we derived between $R_{\rm MOA}$ and the OGLE-IV magnitudes 
\citep{gould_col,bennett12} yields
\begin{align}
I_{\rm cal} =& I_{\rm O4} - (0.029 \pm 0.010)(R_{\rm MOA}-I_{\rm O4}) - 0.0111 \pm 0.0004
\label{eq-cal_I} \ , \\
V_{\rm cal} =& I_{\rm O4} + (4.845 \pm 0.061)(R_{\rm MOA}-I_{\rm O4}) + 0.1749 \pm 0.0010
\label{eq-cal_V} \ ,
\end{align}
for the calibrated $I$ and $V$ magnitudes in terms of the $R_{\rm MOA}$ and $I_{\rm O4}$ magnitudes
from the light curve models. The zero point of the $R_{\rm MOA}$ magnitude system used in this 
paper is 28.1415, which is designed to give a color of $R_{\rm MOA}-I_{\rm O4}$ 
when $V_{\rm O4}-I_{\rm O4}= 0$.
With these calibration relations we find the source magnitudes given in Table~\ref{tab-mparams}, namely
$I_s = 20.747$, $V_s = 23.475$ for the best fit model and
$I_s = 20.747\pm 0.035$ and $V_s = 23.468\pm 0.071$ for the average of our MCMC runs. 
Figure~\ref{fig-cmd} shows the calibrated OGLE-IV color magnitude diagram in black. 
The \citet{holtzman98} Hubble Space Telescope (HST) CMD for Baade's window shifted to the same
extinction and bar distance as the MOA-2011-BLG-291 is plotted in green. The source star is 
indicated in blue, and it is clearly redder or brighter than the bulge main sequence
of \citet{holtzman98}. 
While there are some stars in this region of the CMD, \citet{clarkson08}
have shown that the stars in this region of the CMD are almost entirely low mass main sequence 
stars in the foreground disk.
The MOA-2012-BLG-291 field should have many more of these than the \citet{holtzman98} field
used for the bulge CMD, because the the MOA-2011-BLG-291 field is about a factor of 2 closer
to the Galactic plane than the Baade's window field. Thus, this
source star could be located in the foreground disk.

In order to estimate the source radius, we need extinction-corrected magnitudes,
which can be determined from the magnitude and color of the centroid of the
red clump giant feature in the CMD, as indicated
in Figure~\ref{fig-cmd} \citep{yoo_rad}. We find that the
red clump centroid in this field is at $I_{\rm cl} = 16.005$, $(V-I)_{\rm cl} = 2.428$, 
which implies $V_{\rm cl} = 18.433$. From \citet{nataf13}, we find that the extinction corrected
red clump centroid should be at $I_{\rm cl,0} = 14.401$, $(V-I)_{\rm cl0,} = 1.06$, which implies
$I$ and $V$-band extinctions of $A_I = 1.604$ and $A_V = 2.972$. So, the extinction corrected
source magnitude and color are $I_{s0} = 19.143$ and $(V-I)_{s0} = 1.353$ for the best fit model.
These dereddened magnitudes can be used to determine the angular source radius,
$\theta_*$. With the source magnitudes that we have measured, the most precise determination
of $\theta_*$ comes from the $(V-I),I$ relation. We use
\begin{equation}
\log_{10}\left[2\theta_*/(1 {\rm mas})\right] = 0.501414 + 0.419685\,(V-I)_{s0} -0.2\,I_{s0} \ ,
\label{eq-thetaS}
\end{equation}
which comes from the \citet{boyajian14} analysis, but with the color range optimized for the 
needs of microlensing surveys. These numbers are not included
in the \citet{boyajian14} paper, but they were provided in a private communication from 
T.S.\ Boyajian (2014). There are three effects that influence the uncertainty in the
angular source radius, $\theta_*$. These are the intrinsic uncertainty in the source magnitude
and color, the uncertainty in the angular radius relation (equation~\ref{eq-thetaS}), and the
uncertainty in the extinction. There is a partial cancelation of the uncertainties due to 
extinction. An increase in extinction will make the extinction corrected source magnitude brighter,
which would increase $\theta_*$, but it would also make the extinction corrected color bluer, which 
would decrease $\theta_*$. The uncertainty in the source magnitude and color are correlated
with the uncertainty in $t_E$, due to the blending degeneracy \citep{yee12}. 
This blending degeneracy occurs 
because a light curve with a fainter, smaller source, smaller $u_0$, and 
larger $t_E$ has a close resemblance to the original light curve. This correlation is
important for the determination of $\theta_E = \theta_* t_E/t_*$. Therefore, we handle this 
uncertainty in our MCMC calculations, so as to include all the correlations in our determination
of the lens system properties.
For the best fit model parameters listed in Table~\ref{tab-mparams}, we find 
$\theta_* = 0.882\pm 0.054\,$mas, where the uncertainty includes only the uncertainties in the 
extinction and the source radius relation, equation~\ref{eq-thetaS}.

\section{Lens System Properties}
\label{sec-lens_prop}

As discussed in Section~\ref{sec-radius}, the angular Einstein radius, 
$\theta_E = \theta_* t_E/t_*$, can be determined from light curve parameters,
as long as the angular source size, $\theta_*$, can be determined from the 
source brightness and color. The determination of $\theta_E$ allows us to use
the following relation \citep{bennett_rev,gaudi_araa}
\begin{equation}
M_L = {c^2\over 4G} \theta_E^2 {D_S D_L\over D_S - D_L} 
%       =  {c^2\over 4G} \theta_E^2 {{\rm AU}\over \pi_{\rm rel}}
       = 0.9823\,\msun \left({\theta_E\over 1\,{\rm mas}}\right)^2\left({x\over 1-x}\right)
       \left({D_S\over 8\,{\rm kpc}}\right) \ ,
\label{eq-m_thetaE}
\end{equation}
where $x = D_L/D_S$. This expression is often considered to be a mass-distance relation. However,
in the case of MOA-2011-BLG-291, the source does not lie on the main sequence of Galactic bulge
stars, so we consider several possible constraints on the source distance.

In order to construct a prior to constrain the source brightness as a function of distance, we have
two options: theory or observations, and we will consider both options.  For the empirical relations,
we use the
same empirical mass-luminosity relation that was used in \citet{bennett15,bennett16}. We use the mass-luminsity
relations of \citet{henry93}, \citet{henry99} and \citet{delfosse00} in different mass
ranges. For $M_L > 0.66\,\msun$, we use the \citet{henry93} relation; for
$0.12\,\msun < M_L < 0.54\,\msun$, we use the \citet{delfosse00} relation; and for
$0.07 \,\msun < M_L < 0.10\,\msun$, we use the \citet{henry99} relation. In between these
mass ranges, we linearly interpolate between the two relations used on the
boundaries. That is, we interpolate between the \citet{henry93} and the \citet{delfosse00}
relations for $0.54\,\msun < M_L < 0.66\,\msun$, and we interpolate between the
\citet{delfosse00} and \citet{henry99} relations for $0.10\,\msun < M_L < 0.12\,\msun$.
These relations only provide magnitudes in the $V$, $J$, $H$, and $K$ passbands, so to obtain
relations for the $I$-band, we use the color transformations presented in \citet{kenyon95}.
We have also checked the more recent analysis of \citet{benedict16} to replace the low-mass relations
of \citet{henry99} and \citet{delfosse00}, and the results change very little.

\begin{figure}
\epsscale{0.9}
\plotone{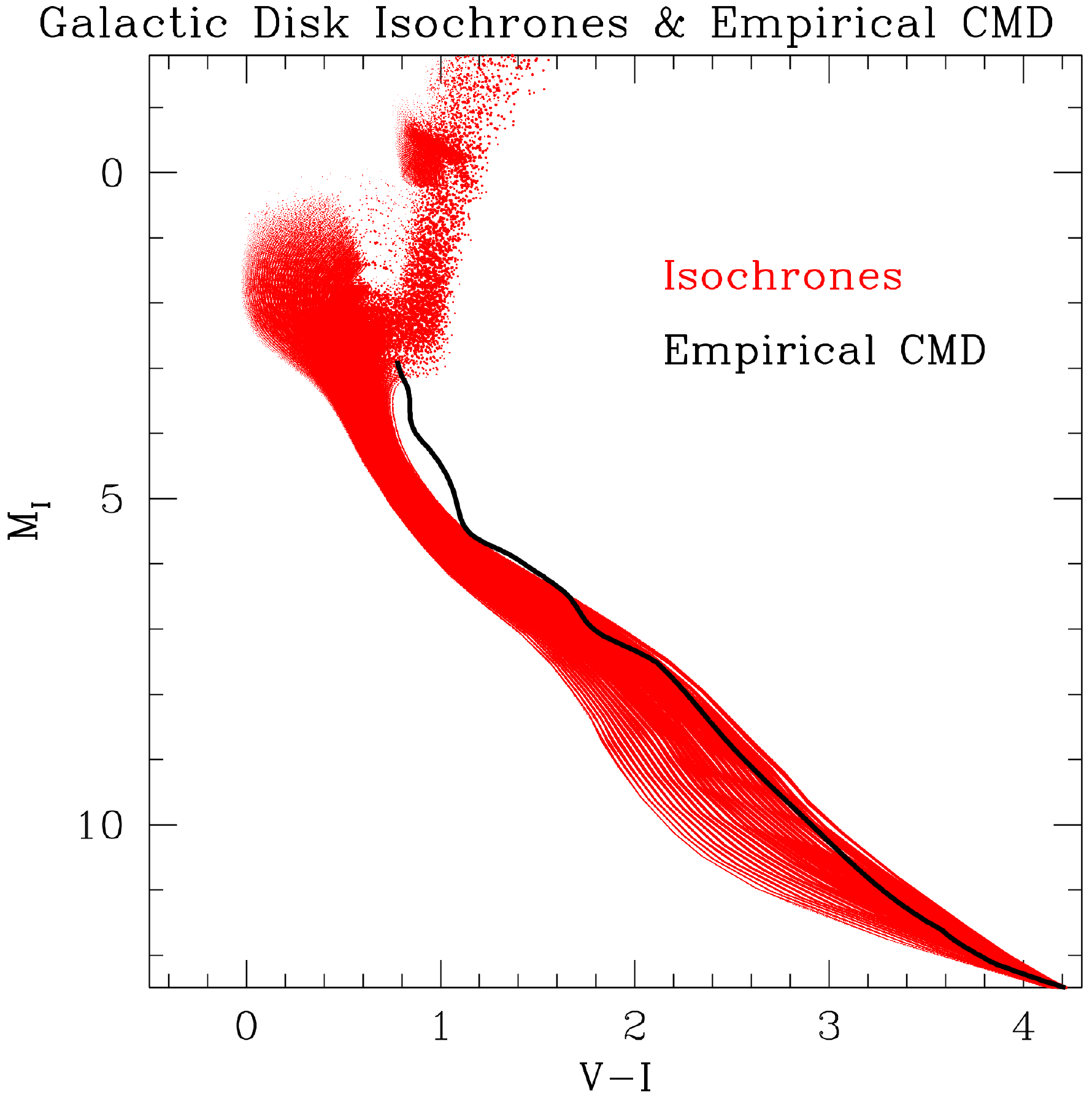}
\caption{The $(V-I,I)$ CMD constructed from Galactic disk isochrones (in red) and
a CMD constructed from empirical mass-luminosity relations (black curve). 
\label{fig-cmd_disk_iso}}
\end{figure}

\begin{figure}
\epsscale{0.9}
\plotone{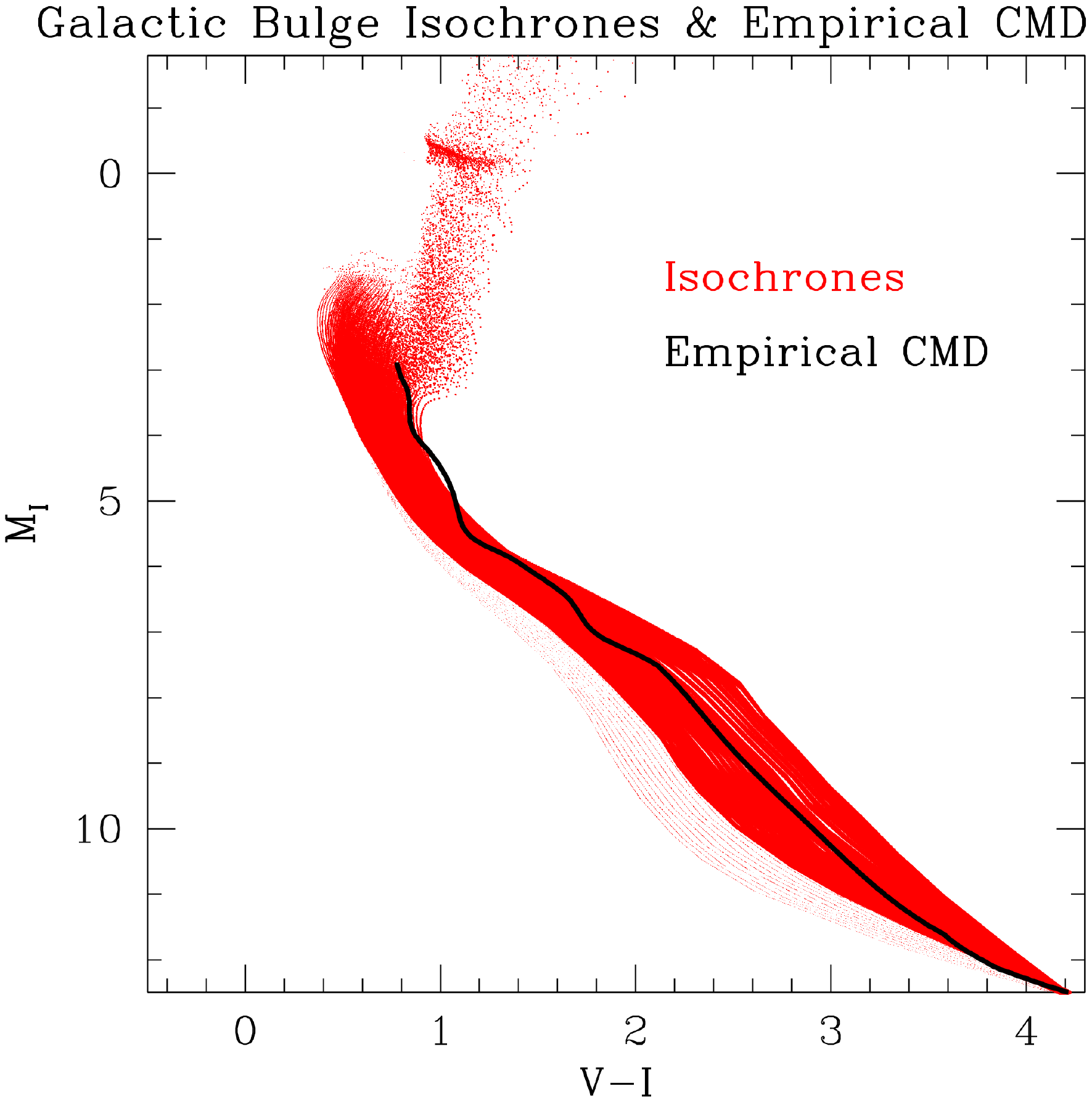}
\caption{The $(V-I,I)$ CMD constructed from Galactic bulge isochrones (in red) and
a CMD constructed from empirical mass-luminosity relations (black curve). 
\label{fig-cmd_blg_iso}}
\end{figure}

For the theoretical mass-luminosity relations, we
use isochrones from the PAdova and tRieste Stellar Evolution Code (PARSEC) project
\citep{bressan12_PARSEC,chen14_PARSEC,chen15_PARSEC,tang14_PARSEC}.
In order to avoid biasing our results with an overly restrictive prior, we chose a wide range of 
ages and metalicities for our prior. The isochrone grids available from PARSEC are
spaced logarithmically in age from $\log({\rm Age}/1 {\rm yr}) = 8.8$ to $\log({\rm Age}/1\,{\rm yr}) = 10.1$
at an interval of 0.05, and the metalicity intervals are spaced approximately logarithmically with intervals
of $0.05\,$dex or $0.06\,$dex. For our Galactic disk isochrone priors, we use ages between
$1\,$Gyr and $10\,$Gyr, with a weighting of 1 for $3.8\,{\rm Gyr} < {\rm Age} < 6.7\,{\rm Gyr}$. For younger
ages, the weights decrease linearly down to a weight of 0.1 at  ${\rm Age} = 1\,$Gyr, and for older ages
the weights decrease linearly down to a weight of 0.5 at  ${\rm Age} = 10\,$Gyr. For the disk isochrones, we
use metalicities between $\log Z = -2.8$ and $\log Z = -1.3$. Isochrones with metalicities 
$-2.26 < \log Z < -1.93 $ are given unit weight, while the weights decrease linearly from
$\log Z = -2.30$ down to $\log Z = -2.80$ and from $\log Z = -1.94$ up to $\log Z = -1.30$.
These isochrones are compared to our empirical mass-luminosity relation in Figure~\ref{fig-cmd_disk_iso}.

Our bulge isochrones primarily cover the metalicity range $-2.3 \leq \log Z \leq -1.3 $ with ages in the 
$2.0\,{\rm Gyr} \leq {\rm Age} \leq 12.6\,{\rm Gyr} $ range with uniform weights in $\log Z$ and Age.
In addition, we also include a contribution from old ($10.0\,{\rm Gyr} \leq {\rm Age} \leq 12.6\,{\rm Gyr} $,
low metalicity ($-2.8 \leq \log Z < -2.3 $) stars with a weight that is 7\% of the weight of the higher metalicity
stars of the same age. Figure~\ref{fig-cmd_blg_iso} compares these isochrones to our empirical 
mass-luminosity relation. For our selection of disk isochrones, the empirical relation is generally brighter
or redder than the isochrones, particularly for $V-I < 1.1$. The agreement between the empirical relation and
the isochrones is a bit better for our selection of bulge isochrones, but the discrepancy at $V-I < 1.1$ remains.
Of course, the empirical relation is based on stars in the disk, so it is the comparison with the disk isochrones
that is a more reasonable test. This discrepancy at $V-I < 1.1$ seems to be primarily a problem 
with the \citet{henry93} relations, which provides colors that seem far too red for stars in the 
$0.7\msun < M \leq 1.0\msun$ mass range. However, at lower masses, the empirical relations seem more
reliable as recent studies \citep{benedict16} give similar results to studies that are almost two decades
old \citep{henry99,delfosse00}. 

For masses $> 0.9\msun$, the isochrones have an additional advantage. Stars in this mass range may
have exhausted the Hydrogen in their cores, so they may have evolved to reach the main sequence 
turn-off or the giant branch. An example of this is the  host star for the two planet event, 
OGLE-2012-BLG-0026 has been shown to be a $1.06\pm 0.05\msun$ turn-off star \citep{beaulieu16},
and another example is the stellar binary source system for planetary event, MOA-2010-BLG-117
\citep{bennett18}. Both sources are subgiants. Thus, it appears that it is probably best to use the 
empirical relations for stars with mass $< 0.7\msun$ and isochrones for stars that have masses
$> 0.9\msun$.

\subsection{Bayesian Analysis with Source Magnitude and Color Constraints}
\label{sec-bayes_constraint)}

\begin{deluxetable}{ccccccc}
\tablecaption{Physical Parameters\label{tab-pparam}}
\tablewidth{0pt}
\tablehead{
\colhead{Parameter}  & \colhead{units} & \colhead{No source} & \colhead{Empirical} &  \colhead{2-$\sigma$ range} &  \colhead{Isochrone} &  \colhead{2-$\sigma$ range} \\
& & constraint & & & & } 
\startdata 
$\theta_E$ & mas & $0.140\pm 0.012$ & $0.129\pm 0.012$ & 0.108-0.155  & $0.123\pm 0.011$ & 0.104-0.147 \\
$\mu_{\rm rel,G}$ & mas/yr & $2.13\pm 0.19$ & $2.01\pm 0.18$ & 1.69-2.40 & $1.92\pm 0.16$ & 1.62-2.27 \\
$D_S $ & kpc & $7.9\pm 1.7$ & $4.9\pm 1.3$ & 3.3-9.1 & $3.8\pm 0.6$ & 2.6-5.1 \\
$D_L $ & kpc & $7.0\pm 1.5$ & $4.4\pm 1.4$ & 2.7-8.4 & $3.5\pm 0.6$ & 2.4-4.7 \\
$M_\star$ & $\msun$ & $0.22^{+0.32}_{-0.13}$ & $0.15^{+0.27}_{-0.10}$ & 0.02-0.72 & $0.13^{+0.25}_{-0.08}$ & 0.02-0.64  \\
$m_p$ & $\mearth$ & $27^{+41}_{-17}$ & $18^{+34}_{-12}$ & 2-94  & $16^{+31}_{-11}$ & 2-85 \\
$a_\perp$ & AU & $1.2\pm 0.3$ & $0.69\pm 0.24$ & 0.38-0.82  & $0.52\pm 0.11$ & 0.32-0.76 \\
$a_{3d}$ & AU & $1.4^{+0.8}_{-0.3}$ & $0.77^{+0.59}_{-0.20}$ & 0.42-2.40 & $0.62^{+0.32}_{-0.17}$ & 0.35-1.80\\
%$P$ & yr & $7.6{+7.7\atop -1.4}$ & 5.4-62 \\
%$V_L$ & mag & $22.81{+0.09\atop -0.07}$ & 22.68-23.05 \\
%$I_L$ & mag & $20.48{+0.08\atop -0.11}$ & 20.28-20.64 \\
%$H_L$ & mag & $18.25\pm 0.11$ & 18.03-18.42 \\
\enddata
\tablecomments{ Mean values and RMS for $\theta_E$, $\mu_{\rm rel,G}$, $D_L $ and $a_\perp$. 
Median values and 68.3\% confidence intervals are given for the rest. 2-$\sigma$ range refers to the
95.3\% confidence interval.}
\end{deluxetable}

The Galactic bulge CMD from Baade's Window that has been transformed to match the centroid of the
red clump giant feature in the field of MOA-2011-BLG-291, shown in Figure~\ref{fig-cmd} indicates 
that the source is brighter or redder than the main sequence. The vicinity of the source star does have
a low density of stars, but the multi-epoch observations of the Galactic bulge SWEEPS field by
\citet{clarkson08} allowed the separation of bulge and foreground disk stars based on their proper 
motion. \citet{clarkson08} showed that the stars just above and redder than the main sequence 
are foreground disk stars. The MOA-2011-BLG-291 field, at a Galactic latitude of $b = -1^\circ.97$
should have many more foreground disk stars than the Baade's Window field of \citet{holtzman98}
at $b = -3^\circ.9$.

\begin{figure}
\epsscale{1.08}
%\plotone{lens_prop_moa291_a3d_LS.pdf}
\plottwo{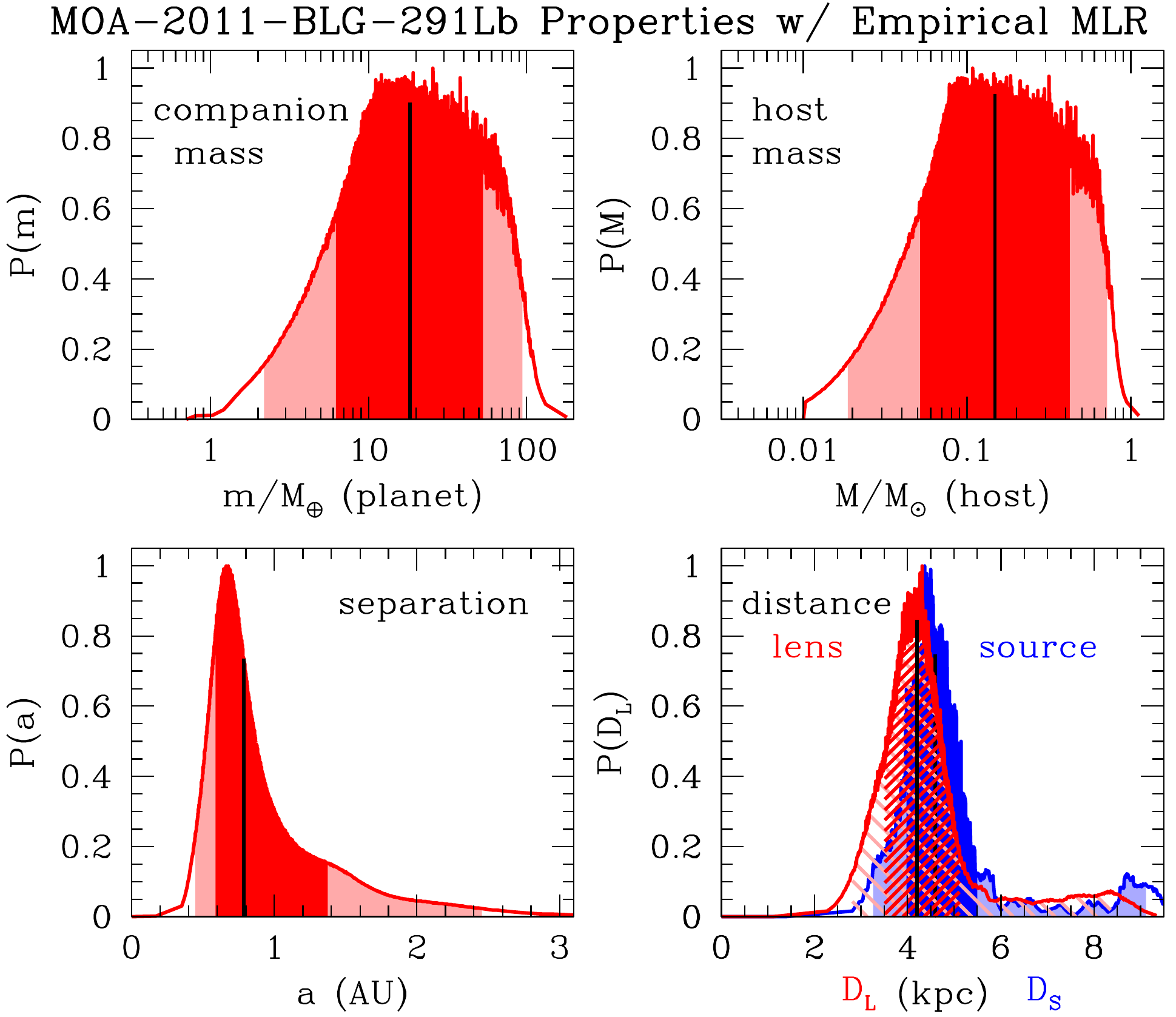}{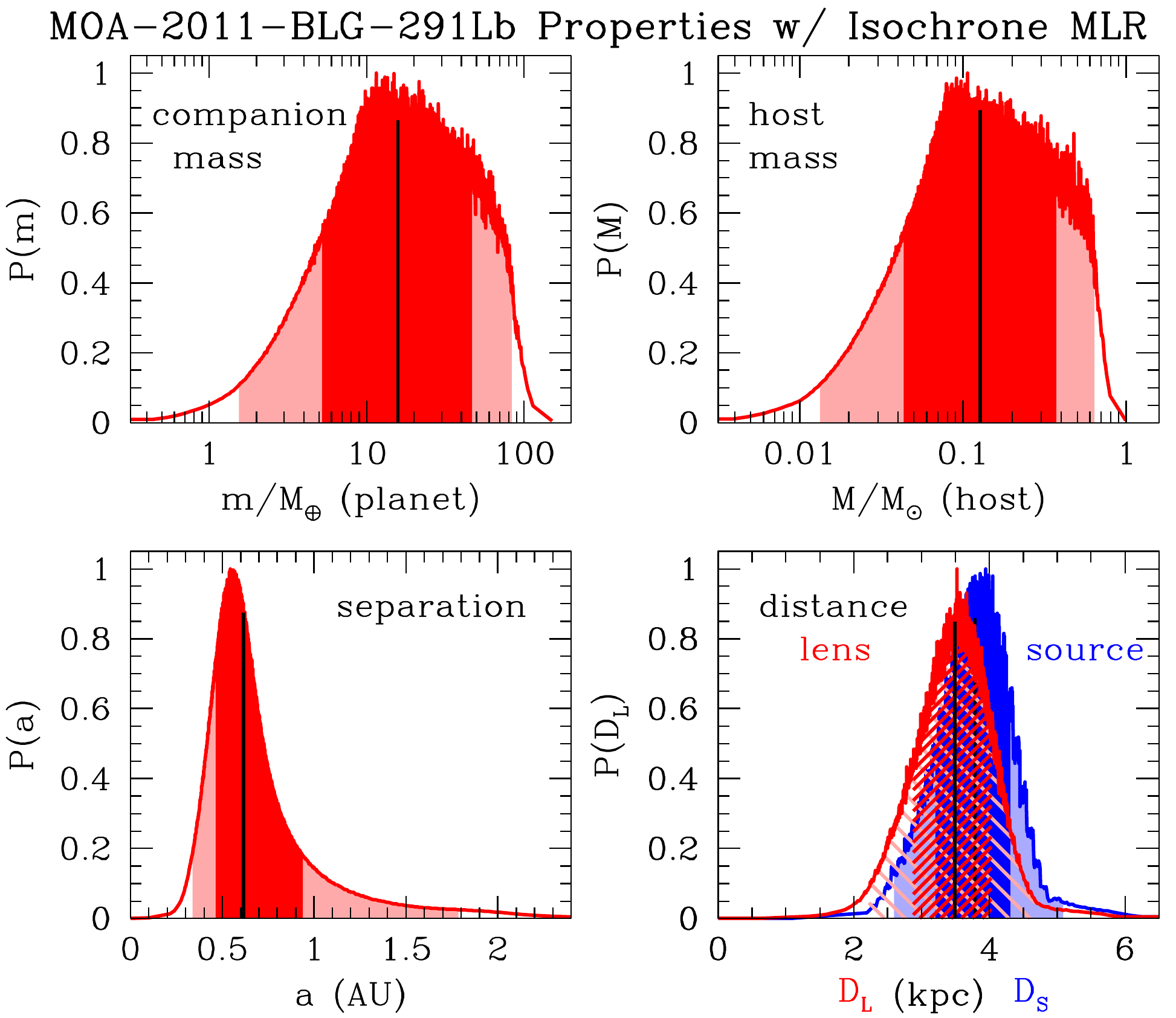}
\caption{The lens properties from our Bayesian analysis with constraints on the source magnitude 
and color from our empirical mass-luminosity in the four left panels, and with our collection of
isochrones in the four right panels. The red histograms represent the lens system masses, separation
and distance, and the blue histogram indicates the source distance.
\label{fig-lens_prop}}
\end{figure}

In order to estimate the lens properties, we perform a Bayesian analysis based on the Galactic model
that includes the lensing probability as a function of the source and lens distances ($D_S$ and $D_L$)
and the lens-source relative proper motion, $\mu_{\rm rel,G}$, in an intertial Geocentric coordinate
system that moves with the Earth at the time of the microlensing light curve peak. The details of the
Galactic model are given in \citet{bennett14}.

The new feature in this analysis is that we demand that source star have a magnitude and color
consistent with our measurements of the source. We use both the empirical relations and the selection
of isochrones described above. The magnitude and color uncertainties due to modeling, listed in 
Table~\ref{tab-mparams}, are accounted for in the Markov Chain calculations. In addition, we include 
a 0.062 mag uncertainty in the transformations, equations~\ref{eq-cal_I}, and \ref{eq-cal_V}, from 
$R_{\rm MOA}-I_{\rm O4}$ to $V_{\rm cal}-I_{\rm cal}$. Finally, we also include mass-luminosity
function uncertainties of $\sigma_{V-I} =  0.05$,  $\sigma_I =  0.15$ for the empirical model and 
$\sigma_{V-I} =  0.05$,  $\sigma_I =  0.10$ for the isochrones. 

Table~\ref{tab-pparam} and  Figure~\ref{fig-lens_prop} show the results of these Bayesian analyses. 
These results indicate that the constraint on the source magnitude and color moves the likely lens
distance from $D_L = 7.0 \pm 1.5\,$kpc to $D_L = 4.4 \pm 1.3\,$kpc and $D_L = 3.5 \pm 0.6\,$kpc 
in the empirical and isochrone constraint cases, respectively. The isochrone constraint implies a 
smaller distance for the lens because the isochrone constraint on the source implies a fainter magnitude 
for a star constrained to have the dereddened color of the source star,
$(V-I)_{s0} = 1.36 \pm 0.08$. This is clear from Figure~\ref{fig-cmd_disk_iso}, and the smaller 
$D_S$ value implies a smaller $D_L$ value. These smaller $D_S$ and $D_L$ values imply
smaller host star ($M_*$) and planet ($m_p$) masses by $\sim 0.5$-$0.9\,\sigma$. The difference between
the values implied by the empirical and isochrone constraints on the host star and planet masses is
much smaller at $\sim 0.2\,\sigma$. The posterior $D_L$ and $D_S$ values for the empirical source
magnitude and color constraint do have a larger tail and even a small peak at Galactic bulge distances
($D_S \simgt 8\,$kpc). This is due to the somewhat larger error bar we have assumed for the 
empirical $I$-band magnitude distribution, combined with the large prior probability for a source star in the
Galactic bulge.

Figure~\ref{fig-lens_prop} clearly shows that the output distributions from our Bayesian analysis are 
quite similar with either the empirical or the isochrone constraints. The preferred masses for the host star and plant
are $M_* \sim 0.15\msun$ and $m_p \sim 18\mearth$, but the range of masses allowed at 95\%
confidence is quite large: a factor of 47 for $m_p$ and a factor of 36 for $M_*$. This is a consequence of
equation~\ref{eq-m_thetaE}, since this indicates that the lens system mass scales as 
$\propto 1/(1-x)$ as $x = D_L/D_S$ approaches 1. The small $\theta_E$ and lens-source
relative proper motion, $\mu_{\rm rel,G}$, values favor a lens close to the source. 
The ``G" suffix in $\mu_{\rm rel,G}$ indicates that the relative proper motion is measured in
a ``Geocentric" inertial frame that instantaneously moves with the orbital motion of the Earth
at the time of the microlensing light curve peak.
The last two parameters in this table are the planet-star separation projected to the plane of the
sky, $a_\perp$, and the three dimensional separation, $a_{3d}$, under the assumption that the orientation 
of the planetary orbit is random. This is a good assumption for planetary systems in general, but 
it is not necessarily a good assumption for a system with a planet discovered by microlensing. If, for
example, planets in very wide orbits are more common than those in orbits of a few AU, then
the discovered planets would tend to have large separations along the line of sight, as this would push
the projected separation to the range of highest sensitivity with the microlensing method.

\subsection{Bayesian Analysis with Excess Extinction}
\label{sec-extinct_constraint)}

An alternative explanation for the unusually red color of the source is that the source could experience 
significantly more dust extinction than the average of the red clump stars that appear in our CMD
(see Figure~\ref{fig-cmd}). This becomes more likely for events close to the Galactic plane, like this event
at Galactic latitude $l = -1.97$, because galactic dust has a small scale height \citep{drimmel}.
However the CMD for this event does not show a pronounced elongation of the red clump along the 
reddening vector, at $(A_I, E(V-I)) = (1.604, 1.368)$. Events at even lower galactic latitudes, like
OGLE-2013-BLG-1761 at $ b = -1.48$ \citep{hirao17} and OGLE-2015-BLG-1670 at $b = -1.12$
\citep{ranc18} are more likely to exhibit this excess extinction effect, and in fact, 
the source for OGLE-2013-BLG-1761 is also anomalously red. For the one microlens planet discovered
with infrared data only, UKIRT-2017-BLG-001Lb \citep{shvartzvald18}, at $b = -0.33$, the extinction of the
source star is almost certainly larger than that of the red clump stars, but this source is likely to be in the
background disk.

\begin{deluxetable}{ccccc}
\tablecaption{Physical Parameters from Isochrone Constraints \label{tab-pparamISO}}
\tablewidth{0pt}
\tablehead{
\colhead{Parameter}  & \colhead{units} &  \colhead{Isochrone} &   \colhead{$\Delta E_{V-I} =0.36$} & \colhead{$\Delta E_{V-I} =0.62$}}
\startdata 
$\theta_E$ & mas & $0.123\pm 0.011$ & $0.124\pm 0.011$ &$0.110\pm 0.009$ \\
$\mu_{\rm rel,G}$ & mas/yr & $1.92\pm 0.16$ & $1.92\pm 0.16$  & $1.72\pm 0.13$ \\
$D_S$ & kpc & $3.8\pm 0.6$ & $10.7\pm 1.5$ & $9.7\pm 0.5$ \\
$D_L $ & kpc & $3.5\pm 0.6$ & $8.8\pm 1.5$ & $8.6\pm 0.9$ \\
$M_\star$ & $\msun$ & $0.13^{+0.25}_{-0.08}$ & $0.11^{+0.28}_{-0.06}$ & $0.15^{+0.30}_{-0.09}$ \\
$m_p$ & $\mearth$ & $16^{+31}_{-11}$ & $13^{+24}_{-7}$ & $18^{+38}_{-11}$\\
$a_\perp$ & AU & $0.52\pm 0.11$ & $1.3\pm 0.2$ &  $1.13\pm 0.15$ \\
$a_{3d}$ & AU & $0.62^{+0.32}_{-0.17}$ & $1.6^{+0.8}_{-0.3}$ & $1.3^{+0.7}_{-0.2}$ \\
%$P$ & yr & $7.6{+7.7\atop -1.4}$ & 5.4-62 \\
%$V_L$ & mag & $22.81{+0.09\atop -0.07}$ & 22.68-23.05 \\
%$I_L$ & mag & $20.48{+0.08\atop -0.11}$ & 20.28-20.64 \\
%$H_L$ & mag & $18.25\pm 0.11$ & 18.03-18.42 \\
\enddata
\tablecomments{ Mean values and RMS for $\theta_E$, $\mu_{\rm rel,G}$, $D_L $ and $a_\perp$. 
Median values and 68.3\% confidence intervals are given for the rest. }
\end{deluxetable}

\begin{figure}
%\epsscale{0.7}
\epsscale{1.08}
\plottwo{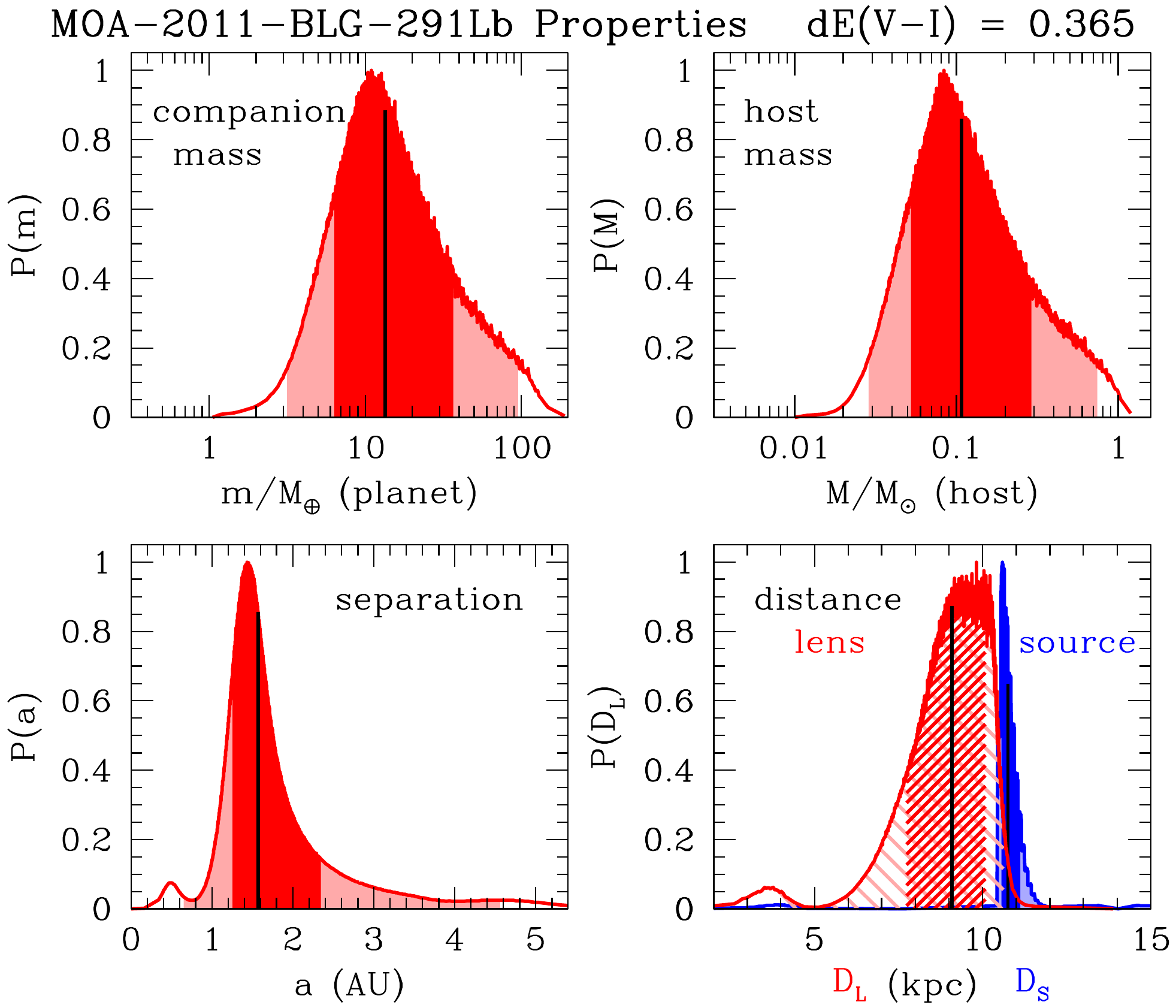}{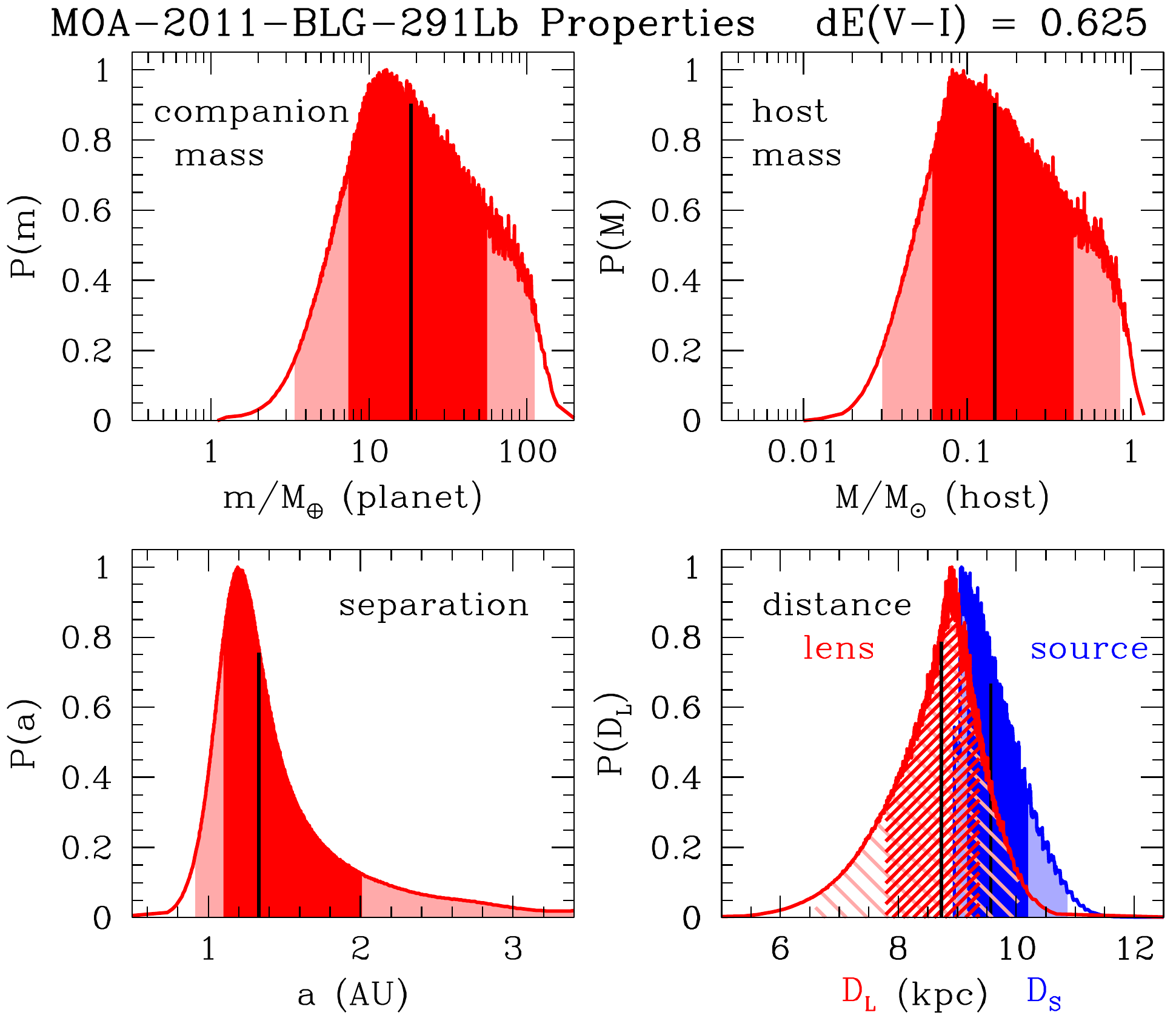}
\caption{Lens properties, as in Figure~\ref{fig-lens_prop}, with models that attempt to explain the 
unusually red source with excess
dust extinction instead of a source in the foreground of the bulge. The 4 plots on the left assume 
a reddening excess of $dE(V-I) = 0.365$ at $D_{\rm dust} = 10.5\,$kpc to enable source stars on the
subgiant branch.  The 4 right panels assume excess extinction of $dE(V-I) = 0.625$ at 
$D_{\rm dust} = 9.0\,$kpc, which was selected to allow upper main sequence and turn-off 
star sources.
\label{fig-lens_prop_TOSG}}
\end{figure}

Although the source seems most likely to be in the foreground disk, let us consider the alternative possibility
that the source could suffer excess extinction. If this excess extinction were removed, it would cause the
source to move toward the upper left in the CMD, as indicated by the solid and short-dashed arrows in
Figure~\ref{fig-cmd}. The source star must be at a greater distance than the bulk of the bulge main 
sequence and giant branch stars to have higher extinction, so the source stars with excess extinction
should well beyond the distance of the Galactic center. We consider two possibilities: a source near the
top of the main sequence and a source on the giant branch. We need
excess extinction of roughly $dE(V-I) = 0.365$ to put the source on the giant branch, and we put this 
extinction at the far side of the bulge at $D_{\rm excess} = 10.5\,$kpc. This is $> 2\,$kpc beyond the 
center of the Galaxy, and might plausibly be where the far side of the Galactic bar merges with the
inner boundary or the far side of the Galactic disk. So, it seems possible that there could be some
excess dust at this location, although our choice of a highly localized cloud just at this distance
is chosen for convenience and to explain our result.

For a source at the top of the main sequence or on the main sequence turn-off, we need excess
reddening of $dE(V-I) = 0.625$, but the excess extinction also makes the source fainter, and if
we put this excess extinction at a distance much beyond the center of the Galaxy, then the 
measured brightness of the source will be too large to be consistent with a main sequence G-dwarf.
To avoid this problem, we add the excess extinction at $D_{\rm excess} = 9.0\,$kpc, although that seems 
physically less plausible than the smaller amount of excess dust we added at $D_{\rm excess} = 10.5\,$kpc.

The results of the Bayesian analyses for sources assumed to have higher extinction than the 
average extinction of the red clump giants are shown in Figure~\ref{fig-lens_prop_TOSG} and compared
to the Bayesian results for no excess extinction in Table~\ref{tab-pparamISO}. We consider only
the isochrone mass-luminosity relations for this comparison because they are the only relations
that cover the evolved source stars that are favored with excess extinction. The most striking thing
about this comparison is that the different distances to the source and lens have little effect on
the likely lens system masses. The host and planet masses for the sources with excess
extinction are within $0.5\sigma$ of the lens masses for the case of the foreground disk sources. 
This is largely a consequence of the relatively small lens-source relative proper motion, 
$\mu_{\rm rel} \sim 2\,$mas/yr. This provides a relatively low probability for having the source and lens
in different stellar populations, like the bulge and disk, and it helps to ensure that the liens is
likely to be located very close to the source. The only exception to this rule is for sub-giant sources
located in the far side of the disk. Their disk orbital motion, counter to the direction of the Sun's motion,
gives the source a large proper motion of $\sim 8.5\,$mas/yr, which is much larger than the 
velocity dispersion of the disk stars, which share the Earth's orbital motion about the Galactic 
center until the distance to the lens drops to a few kpc. This is the reason for the small bump
in the probability distribution at $D_L \sim 3\,$kpc for $dE(V-I) = 0.365$ in Figure~\ref{fig-lens_prop_TOSG}.

The only significant differences in the predicted planetary system properties between the excess extinction 
scenario and the foreground disk source scenario
is the distance to the planetary system, $D_L$, and the planet-star separation, which is proportional
to $D_L$ (because $\theta_E$ is constrained by the light curve, with a modest variation
due to the different extinction values). The excess extinction
models predict planetary system distances and planet-star separations that are a little more than a factor of
two larger than the models with no excess extinction. This change in $D_L$ with little change in the 
host star mass, $M_*$, implies a significant shift in the lens star magnitude. We find $K$-band magnitudes
of $K_L = 23.4{+1.8\atop -2.0}$ without excess extinction, but $K_L = 25.5{+8.0\atop -2.0}$ with
$dE(V-I) = 0.365$ and $K_L = 24.7{+8.6\atop -2.2}$ with $dE(V-I) = 0.625$. Because of the broad
range of masses allowed for the lens, there is a large overlap in these magnitude ranges, so detection
of the lens star will not help to determine whether either of these excess extinction scenarios are correct,
unless there are additional constraints on the source distance, $D_S$.
A measurement of the source proper motion, could help to distinguish these scenarios as the 
proper motion of stars on the near and far side of the disk are quite different. The
velocity dispersion of the bulge is large enough so that a proper motion measurement may not
yield a definitive location of the source star, but non-definitive constraints can still be very useful
in the context of a statistical analysis with a large sample of exoplanets.

\section{Discussion and Conclusions}
\label{sec-conclude}

We have presented the light curve analysis for planetary microlensing event MOA-2011-BLG-291, 
which reveals a planet with a mass ratio of $q = (3.8\pm 0.7)\times 10^{-4}$. The source star for this
event is redder and/or brighter than expected for a star on the Galactic bulge main sequence. There are
several other low latitude planetary microlensing events with sources that appear to be redder or
slightly redder than the normal bulge main sequence and giant branch. There are two obvious mechanisms
that might cause a low latitude source to be unusually red. The star could suffer more dust extinction
than the typical bulge star along the line of site. The dust is expected to have a lower scale height 
than the stars \citep{drimmel}, and it is possible that there is some dust in the bulge. This might be
a better explanation for stars that are on the red edge of the normal bulge populations, such as
OGLE-2015-BLG-1670 \citep{ranc18}, OGLE-2016-BLG-0596 \citep{mroz17_ob160596}, and
OGLE-2017-BLG-0173 \citep{hwang18}. The other explanation that applies to stars that are below
the giant branch is that the source is a fainter main sequence star that lies in the foreground of the bulge.
This implies that it should lie above the main sequence, which also implies that it is redder than the
main sequence because fainter main sequence stars are redder.
In the case of MOA-2011-BLG-291, this seems to be the most likely
explanation because of the relatively large separation between the source star position on the CMD
(Figure~\ref{fig-cmd}) and the main sequence. 
 
If we assume that the source star does not experience any excess extinction
beyond that of our extinction model, then we can constrain its 
distance by comparing to empirical mass-luminosity relations or theoretical isochrone calculations. 
The extinction corrected source color of $(V-I)_{s0} = 1.35\pm 0.07$ is in the range where we think that
the relations are more accurate, so we use these relations to give a source distance of 
$D_S = 4.9 \pm 1.3\,$kpc. A Bayesian analysis of the lens system properties gives a lens distance of 
$D_L = 4.4\pm 1.4\,$kpc and host star and planet masses of $M_{\rm host} = 0.15^{+0.27}_{-0.10}\msun$
and $m_p = 18^{+34}_{-12}\mearth$, respectively, as shown in Figure~\ref{fig-lens_prop} 
and Table~\ref{tab-pparam}. The results using the theoretical isochrones instead
of empirical mass luminosity relations yield source and lens distances that are slightly smaller, but
otherwise the implied parameters of the star and planet lens system are quite similar.

We also explore the possibility that the source is redder than the bulge main sequence due to excess
extinction compared the extinction of the bulge red clump population. We consider two possibilities, excess
reddening of $dE(V-I) = 0.365$, which could put the source on the giant branch, and 
$dE(V-I) = 0.625$, which could put the source on the main sequence. The distribution of red clump stars
in the CMD (Figure~\ref{fig-cmd}) does not indicate a great deal of differential extinction, so we assume
that the excess dust is at $D \geq 9\,$kpc. However, as indicated in Figure~\ref{fig-lens_prop_TOSG} and
Table~\ref{tab-pparamISO}, the Bayesian analysis yields very similar lens properties in both 
these excess extinction scenarios and the case no excess extinction, which requires a foreground 
disk source. 

% Dstar, fmu, Vhat =    4900.000       2.100000       48.77810
%  maximum at 0.826    0.48001E-02
% Dstar, fmu, Vhat =    8300.000       1.900000       74.75518
%  maximum at  0.888   0.16563E-01  
% ratio = 0.28980860955141   
% density of disk at 4.9 kpc = 0.144925
% density of bulge at 8.3 kpc = 1.65750
% ratio = 0.087435897435897
% ratio of rate*density = 0.025339675860777 
% lower mass stars are 1.321285140562249 times more common on the MS 
% total factor 0.033480937181509

The lensing rate of disk stars (per unit area) is $\sim 30$ times lower than the rate for bulge sources,
at this line of sight, so a foreground disk source star is not likely for any given event, but given the $\sim 60$
microlens exoplanets already discovered, we would expect that some would have foreground disk source
stars. With a source galactic latitude of $b = -1.9693$, the line of sight will reach $300\,$pc below the Galactic
plane when it reaches a distance of $9\,$kpc. 
Locally, the dust scale height is thought to be $\sim 120\,$pc \citep{drimmel}, and this scale height
is expected to decrease interior to the Solar Circle. Thus, it seems unlikely that either of these excess 
extinction values could be correct, so we expect that it is most likely that the source is located in the 
inner Galactic disk, in the foreground of the bulge.

One test that might help to determine the location of the source star would be to obtain high angular
resolution images of the source and possible lens star with the Hubble Space Telescope (HST)
\citep{bennett06,bennett15,aparna17}, or ground-based adaptive (AO) systems, such as those on the 
Keck telescope \citep{batista14,batista15,beaulieu16}. Multiple epochs of high resolution imaging should reveal
the source proper motion with respect to the other stars in field. However, it is important to be able to 
distinguish between the motions of the source and lens stars, since they are likely to be blended 
with each other. Due to the low lens-source relative 
proper motion of $\mu_{\rm rel,G} = 2.01\pm 0.18$, the lens and source will have overlapping PSFs
for another $\sim 20$ years, and it is possible for excess stellar flux blended with the source to be due to
a star other than the lens star \citep{kosh17_mb16227}. Fortunately, detailed modeling of the blended images
\citep{aparna17} can reveal the true proper motion of the source. Such multi-star modeling can also determine
if the blend star has a $\mu_{\rm rel,G}$ value consistent with the lens-source pair. 

An additional method to determine if excess stellar flux blended with the source is due to the lens and planetary
host star is the color dependent centroid shift method. 
This was used with HST data to confirm the identification of the 
host star for the first exoplanet found by microlensing \citep{bond04,bennett06}. More recently 
\citet{lu14,lu16} have shown that Keck AO imaging can obtain stellar astrometry measurements with 
uncertainties as low as $\sigma_\mu \approx 0.2\,$mas which is slightly better than the precision
that is typically obtained with well designed HST programs in fields as crowded as the Galactic
bulge. As a result, it is possible to measure much larger and higher S/N
color dependent centroid shifts with
near simultaneous optical HST and infrared Keck AO images. This is particularly useful for 
events, like MOA-2011-BLG-291, with low $\mu_{\rm rel,G}$ values that imply low lens-source separations.

Unfortunately, there is a good chance that the measurement of the source proper motion alone will 
not be sufficient to determine if the source star resides in the foreground disk, bulge, or background
disk behind the bulge, because the relative proper motion distributions for these populations overlap. 
However, if the lens brightness can be determined in more than one passband, then the mass-luminosity
relations for these multiple passbands can be combined with the mass-distance relation 
(equation~\ref{eq-m_thetaE}) to yield independent lens mass-magnitude relations for each passband. 
With lens magnitude measurements in two different passbands, we can, in principle, solve for $D_S$
in equation~\ref{eq-m_thetaE}, and with measurements in 3 passbands, we would have some redundancy.
In addition, these same high angular resolution observations should also reveal some other indications of 
the higher extinction background population if this is responsible for the unusually red color of the source
star.

The current design of the WFIRST exoplanet microlensing survey has fields covering the range
in Galactic coordinates of $ -0.5^\circ \simlt l \simlt 1.7^\circ$ and  $-2^\circ \simlt b \simlt -0.4^\circ$
with 5 of the 7 fields located at $-2^\circ \simlt b \simlt -1.2^\circ$ \citep{penny18}. 
Event MOA-2011-BLG-291 is located in the currently planned WFIRST exoplanet microlensing
survey footprint, but it is on the edge of this footprint furthest from the Galactic plane. This event has
presented an unusually red source that might be interpreted as a low-mass foreground disk star or
a brighter background disk star. In this case, we favor the foreground disk source interpretation, but 
this uncertain source distance issue will become more acute at lower $|b|$. This has been clearly
demonstrated by the first infrared-only planetary microlensing event \citep{shvartzvald18}. For that
event, the source is expected to reside on the far side of the disk, beyond the Galactic bulge, but
the source distance, $D_S$, is quite uncertain. Also, the extinction is so high that the source can
only be detected in the $H$ and $K$ passbands. This rules out many of the observations that 
we have discussed above to potentially remove the source distance ambiguity. 

The challenge of determining the lens and source distances is an important issue for the WFIRST
exoplanet microlensing survey \citep{bennett18_wfirst,penny18} because the ability to determine
the exoplanet host masses is an important feature of a space-based microlensing survey
\citep{bennett02,bennett07}. This likely led to the selection of the microlensing survey proposed
for the Microlensing Planet Finder mission \citep{bennett_MPF} to be included in the WFIRST
mission \citep{WFIRST_AFTA}. The lens identification and mass measurements methods 
proposed for WFIRST, have been found to work quite well for events at higher $|b|$ 
\citep{batista15,bennett15,aparna17}, but there is a strong temptation to locate the WFIRST fields
at lower $|b|$ because the event rate is higher there \citep{bennett02,penny18}. So, the selection of
the optimal WFIRST fields will be a balance between the higher microlensing rate at lower $|b|$ and
the difficulty in determining the host star masses at low $|b|$. Thus, attempts to determine the 
the host star masses with follow-up high angular resolution imaging for events, like MOA-2011-BLG-291,
that reside in the candidate WFIRST fields will provide important input information for the 
design of the WFIRST microlensing survey. The other events without giant source stars
in these candidate fields are (in order of increasing $|b|$)
OGLE-2015-BLG-1670 \citep{ranc18}, MOA-bin-1 \citep{bennett12},
OGLE-2013-BLG-1761 \citep{hirao17}, MOA-2011-BLG-293 \citep{yee12,batista14}, 
OGLE-2013-BLG-0341 \citep{gould14}, OGLE-2015-BLG-0966 \citep{street16}, 
OGLE-2006-BLG-109 \citep{gaudi-ogle109,bennett-ogle109}, and
OGLE-2013-BLG-1721 \citep{mroz17_2pl}. High angular resolution follow-up
observations of these events, plus MOA-2011-BLG-291, are strongly encouraged.

\acknowledgments 
DPB, AB, and CR  were supported by NASA through grant NASA-80NSSC18K0274.
The MOA project is supported in Japan by JSPS KAKENHI Grant Numbers JP17H02871, JSPS24253004, 
JSPS26247023, JSPS23340064, JSPS15H00781, and JP16H06287.
The work by CR was supported by an appointment to the NASA Postdoctoral Program at the 
Goddard Space Flight Center, administered by USRA through a contract with NASA.
NJR is a Royal Society of New Zealand Rutherford Discovery Fellow.
The OGLE Team thanks Profs.\ Marcin Kubiak and Grzegorz Pietrzy{\'n}ski for
their contribution to the OGLE photometric data. The OGLE project has
received funding from the National Science Centre, Poland, grant MAESTRO
2014/14/A/ST9/00121 to AU.
The Wise group was supported by the I-CORE programme
of the Planning and Budgeting Committee and the Israel
Science Foundation, Grant 1829/12. DM and AG acknowledge support by
the US-Israel Binational Science Foundation.
Work  by  CH  was  supported  by  the  grant (2017R1A4A1015178)
of the National Research Foundation of Korea.


\begin{thebibliography}{}
%\bibitem[Alard(1997)]{alard97} Alard, C.\ 1997, \aap, 321, 424 
%\bibitem[Alard \& Lupton (1998)]{ala98}Alard, C. \& Lupton, R.H.\ 1998, \apj, 503, 325
%\bibitem[Albrow et al.(2000)]{albrow-97blg41} Albrow, M.D.\ 2000, \apj, 534, 894
\bibitem[Albrow et al.(2009)]{albrow09} Albrow, M.~D. et al.\ 2009, \mnras, 397, 2099
%\bibitem[Alcock et al.(1995)]{macho-par1}Alcock, C., Allsman, R.~A., Alves, D., et al.~1995, \apjl, 454, L125
%\bibitem[Alcock et al.(1997)]{macho-95b30} Alcock, C., Allen, W.H., Allsman, R.A., et al.~1997, \apj, 491, 436
%\bibitem[An et al.(2007)]{an07} An, D., Terndrup, D.~M., Pinsonneault, M.~H., et al.\ 2007, \apj, 655, 233 
%\bibitem[Anderson \& King (2000)]{andking00} Anderson, J.~\& King, I.~R.\ 2000, \pasp, 112, 1360
%\bibitem[Anderson \& King (2004)]{andking04} Anderson, J.~\& King, I.~R.\ 2004, Hubble Space Telescope
%   Advanced Camera for Surveys Instrument Science Report 04-15
%\bibitem[Bachelet et al.(2012)]{bachelet12} Bachelet, E., Fouqu{\'e}, P., Han, C., et al.\ 2012, \aap, 547, A55
%\bibitem[Baraffe et al.(2005)]{baraffe05} Baraffe, I., et al.\ 2005, \aap, 436, L47
%\bibitem[Barry(2010)]{barry_psf} Barry, R.K., et al.\ 2010,  Proc.\ SPIE, 7731, 77313
%\bibitem[Batalha et al.(2013)]{kepler_16mon} Batalha, N.~M., Rowe, J.~F., Bryson, S.~T., et al.\ 2013, \apjs, 204, 24
%\bibitem[Batista et al.(2011)]{batista11} Batista, V., Gould, A., Dieters, S., et al.\ 2011, \aap, 529, A102
\bibitem[Batista et al.(2014)]{batista14} Batista, V., Beaulieu, J.-P., Gould, A., et al.\ 2014, \apj, 780, 54 
\bibitem[Batista et al.(2015)]{batista15} Batista, V., Beaulieu, J.-P., Bennett, D.P., et al.\ 2015, \apj, 808, 170
%\bibitem[Beaulieu et al.(2006)]{ogle390} Beaulieu, J.-P., Bennett, D.~P., Fouqu{\'e}, P., et al.\ 2006, \nat, 439, 437
\bibitem[Beaulieu et al.(2016)]{beaulieu16} Beaulieu, J.-P., Bennett, D.~P., Batista, V., et al.\ 2016, \apj, 824, 83 
%\bibitem[Beichman et al.(2013)]{beichman13} Beichman, C., Gelino, C.~R., Kirkpatrick, J.~D., et al.\ 2013, \apj, 764, 101 
\bibitem[Benedict et al.(2016)]{benedict16} Benedict, G.~F., Henry, T.~J., Franz, O.~G., et al.\ 2016, \aj, 152, 141\bibitem[Bennett(2008)]{bennett_rev} Bennett, D.P, 2008, in Exoplanets, 
   Edited by John Mason.~Berlin: Springer.~ ISBN: 978-3-540-74007-0,  (arXiv:0902.1761)
\bibitem[Bennett(2010)]{bennett-himag} Bennett, D.P.\ 2010, \apj, 716, 1408
%\bibitem[Bennett et al.(1993)]{bennett-sod} Bennett, D.~P., Alcock, C., Allsman, R., et al.\ 1993, Bulletin of the American Astronomical Society, 25, 1402
\bibitem[Bennett et al.(2018a)]{bennett18_wfirst} Bennett, D.~P., Akeson, R., Anderson, J., et al.\ 2018a, (arXiv:1803.08564)
\bibitem[Bennett et al.(2010a)]{bennett_MPF} Bennett, D.~P., Anderson, J., Beaulieu, J.-P., et al.\ 2010a, arXiv:1012.4486
\bibitem[Bennett et al.(2006)]{bennett06} Bennett, D.~P., Anderson, J., Bond, I.~A., Udalski, A., \& Gould, A.\ 2006, \apjl, 647, L171
\bibitem[Bennett et al.(2007)]{bennett07} Bennett, D.P., Anderson, J., \& Gaudi, B.S.\ 2007, \apj, 660, 781
\bibitem[Bennett et al.(2014)]{bennett14} Bennett, D.~P., Batista, V., Bond, I.~A., et~al.\ 2014, \apj, 785, 155
\bibitem[Bennett et al.(2015)]{bennett15} Bennett, D.~P., Bhattacharya, A., Anderson, J., et al.\ 2015, \apj, 808, 169
%\bibitem[Bennett et al.(2008)]{bennett08}Bennett, D.~P., Bond, I.~A., Udalski, A., et al.\ 2008, \apj, 684, 663
\bibitem[Bennett et al.(2010b)]{bennett-ogle109} Bennett, D.~P., Rhie, S.~H., Nikolaev, S., et~al.\ 2010b, \apj, 713, 837
\bibitem[Bennett \& Rhie(1996)]{bennett96}Bennett, D.P. \& Rhie, S.H.\ 1996, \apj, 472, 660
\bibitem[Bennett \& Rhie(2002)]{bennett02}Bennett, D.P. \& Rhie, S.H.\ 2002, \apj, 574, 985
\bibitem[Bennett et al.(2016)]{bennett16}Bennett, D.P., Rhie, S.H., Udalski, A., et al.\ 2016, \aj, 152, 125
\bibitem[Bennett et al.(2012)]{bennett12} Bennett, D.~P., Sumi, T., Bond, I.~A., et al.\ 2012, \apj, 757, 119 
\bibitem[Bennett et al.(2018b)]{bennett18} Bennett, D.~P., Udalski, A., Han, C., et al.\ 2018b, \aj, 155, 141 
%\bibitem[Bensby et al.(2011)]{bensby11} Bensby, T., Ad{\'e}n, D., Mel{\'e}ndez, J., et al.\ 2011, \aap, 533, A134
\bibitem[Bhattacharya et al.(2017)]{aparna17} Bhattacharya, A., Bennett, D.~P., Anderson, J., et al.\ 2017, \aj, 154, 59 
\bibitem[Bond et al.(2001)]{bond01} Bond, I.~A., Abe, F., Dodd, R.~J., et al.\ 2001, \mnras, 327, 868
\bibitem[Bond et~al.(2004)]{bond04} Bond, I.~A., Udalski, A., Jaroszy{\'n}ski, M.\ 2004,  \apjl, 606, L155
%\bibitem[Bonfils et al.(2011)]{bonfils11} Bonfils, X., Delfosse, X., Udry, S., et al.\ 2011, arXiv:1111.5019
%\bibitem[Borucki et al.(2011)]{borucki11} Borucki, W.~J., Koch, D.~G., Basri, G., et al.\ 2011, \apj, 736, 19
%\bibitem[Boss(1997)]{boss97} Boss, A.~P.\ 1997, Science, 276, 1836
%\bibitem[Boss(2006)]{boss06} Boss, A.~P.\ 2006, \apj, 643, 501
%\bibitem[Boss(2006)]{boss06} Boss, A.P.\ 2006, \apjl, 644, L79
%\bibitem[Bowler et al.(2011)]{bowler11} Bowler, B.~P., Liu, M.~C., Kraus, A.~L., Mann, A.~W., \& 
%Ireland, M.~J.\ 2011, \apj, 743, 148
\bibitem[Boyajian et al.(2014)]{boyajian14} Boyajian, T.S., van Belle, G., \& von Braun, K.,\
   2014, \aj, 147, 47
\bibitem[Bressan et al.(2012)]{bressan12_PARSEC} Bressan, A., Marigo, P., Girardi, L., et al.\ 2012, \mnras, 427, 127 
%\bibitem[Brown et al.(2013)]{lcogt} Brown, T.~M., Baliber, N., Bianco, F.~B., et al.\ 2013, \pasp, in press (arXiv:1305.2437) 
%\bibitem[Butler et al.(2006)]{butler-catalog} Butler, R.~P., Wright, J.~T., Marcy, G.~W., et al.\ 2006, \apj, 646, 505
%\bibitem[Burke et al.(2015)]{burke15} Burke, C.~J., Christiansen, J.~L., Mullally, F., et al.\ 2015, \apj, 809, 8 
%\bibitem[Bramich(2008)]{bramich08} Bramich, D.M.\ 2008, \mnras, 386, L77
%\bibitem[Cardelli et al.(1989)]{cardelli89}  Cardelli, J.A., Clayton, G.C., \& Mathis, J.S.\ 1989, \apj, 345, 245
%\bibitem[Cassan et al.(2012)]{cassan12}Cassan, A., Kubas, D., Beaulieu, J.-P., et al.\ 2012,  \nat, 481, 167
%\bibitem[Carpenter(2001)]{2mass_cal} Carpenter, J.M.\ 2001, \aj 121, 2851
%\bibitem[Claret(2000)]{claret00} Claret, A.\ 2000, \aap, 363, 1081
%\bibitem[Chang \& Refsdal(1979)]{chang-refsdal79} Chang, K., \& Refsdal, S.\ 1979, \nat, 282, 561
%\bibitem[Chang \& Refsdal(1984)]{chang-refsdal84} Chang, K., \& Refsdal, S.\ 1979, \aap, 132, 168
\bibitem[Chen et al.(2015)]{chen15_PARSEC} Chen, Y., Bressan, A., Girardi, L., et al.\ 2015, \mnras, 452, 1068 
\bibitem[Chen et al.(2014)]{chen14_PARSEC} Chen, Y., Girardi, L., Bressan, A., et al.\ 2014, \mnras, 444, 2525 
%\bibitem[Christiansen et al.(2011)]{epoxi} Christiansen, J.L., et al.\ 2011, \apj, 726, 94
\bibitem[Clarkson et al.(2008)]{clarkson08} Clarkson, W., Sahu, K., Anderson, J., et al.\ 2008, \apj, 684, 1110-1142 
%\bibitem[Cumming et al.(2008)]{cumming08}Cumming, A., Butler, R.~P., Marcy, G.~W., Vogt, S.~S., Wright, J.~T., \& Fischer, D.~A.\ 2008, \pasp, 120, 531
%\bibitem[D'Angelo et al.(2010)]{dangelo_book} D'Angelo, G., Durisen, R.~H., \& Lissauer, J.~J.\ 2010,
%   in Exoplanets, ed.\ S.\ Seager (Tucson, AZ: Univ. Arizona Press), 319
\bibitem[Delfosse et al.(2000)]{delfosse00} Delfosse, X., Forveille, T., S{\'e}gransan, D., et al.\ 2000, \aap, 364, 217
%\bibitem[Delorme et al.(2012)]{delhome12} Delorme, P., Gagn{\'e}, J., Malo, L., et al.\ 2012, \aap, 548, A26 
%\bibitem[Di Stefano \& Esin(1995)]{distefano95} Di Stefano, R., \& Esin, A.~A.\ 1995, \apjl, 448, L1
%\bibitem[Di Stefano \& Scalzo(1999)]{distefano99} Di Stefano, R., \& Scalzo, R.~A.\ 1999, \apj, 512, 579
%\bibitem[Dolphin(2000)]{hstphot} Dolphin, A.~E.\ 2000, \pasp, 112, 1383 
%\bibitem[Dong et al.(2006)]{dong06} Dong, S., et al.\ 2006, , \apj, 642, 842
%\bibitem[Dong et al.(2007)]{dong-ogle05smc1} Dong, S., et al.\ 2007, \apj, 664, 862
\bibitem[Dong et al.(2009)]{dong-ogle71} Dong, S., Bond, I.~A., Gould, A., et al.\ 2009, \apj, 698, 1826
%\bibitem[Dong et al.(2009b)]{dong-moa400} Dong, S., Bond, I.~A., Gould, A., et al.\ 2009, \apj, 698, 1826
%\bibitem[Doolin \& Blundell(2011)]{doolin11} Doolin, S., \& Blundell, K.~M.\ 2011, \mnras, 418, 2656
\bibitem[Drimmel \& Spergel(2001)]{drimmel}  Drimmel, R., \& Spergel, D.~N.\ 2001, \apj, 556, 181
%\bibitem[Fruchter \& Hook(2002)]{drizzle} Fruchter A.S.\ 2002, \pasp, 114, 144
%\bibitem[Ford \& Rasio(2008)]{ford08} Ford, E.~B., \& Rasio, F.~A.\ 2008, \apj, 686, 621
\bibitem[Fukui et al.(2015)]{fukui15} Fukui, A., Gould, A., Sumi, T., et al.\ 2015, \apj, 809, 74 
%\bibitem[Furusawa et al.(2013)]{moa328} Furusawa, K., Udalski, A., Sumi, T., et al.\ 2013, \apj, 779, 91
%\bibitem[Gaudi(2010)]{gaudi_rev}Gaudi, B.S.\ 2010, in Exoplanets, ed. S. Seager (Tucson: University of 
%Arizona Press), 79 (arXiv:1002.0332)
\bibitem[Gaudi(2012)]{gaudi_araa} Gaudi, B.~S.\ 2012, \araa, 50, 411
\bibitem[Gaudi et al.(2008)]{gaudi-ogle109} Gaudi, B.~S., Bennett, D.~P., Udalski, A., et al.\ 2008, Science, 319, 927
%\bibitem[Gaudi \& Gould(1997)]{gaudi97} Gaudi, B.S., \& Gould, A.\ 1997, \apj, 486, 85
%\bibitem[Gonzalez et al.(2011)]{vvv_extinct} Gonzalez, O.~A., Rejkuba, M., Zoccali, M., Valenti, E., \& Minniti, D.\ 2011, \aap, 534, A3
%\bibitem[Gould(1992)]{gould-par1} Gould, A.\ 1992, \apj, 392, 442
%\bibitem[Gould(2004)]{gould-jerk} Gould, A.\ 2004, \apj, 606, 319
%\bibitem[Gould(2008)]{gould-hex} Gould, A.\ 2008, \apj, 681, 1593
%\bibitem[Gould(2014)]{gould-1dpar} Gould, A.\ 2014, J.\ Kor.\ Ast.\ Soc., 47, 215
\bibitem[Gould et al.(2010)]{gould_col} Gould, A., Dong, S., Bennett, D.~P., et al.\ 2010a, \apj, 710, 1800
%\bibitem[Gould et al.(2010b)]{gould10}Gould, A., Dong, S., Gaudi, B.S., et al.\ 2010b,  \apj, 720, 1073
%\bibitem[Gould et al.(2004)]{gould04}Gould, A., Gaudi, B.S., \& Han, C., 2004, arXiv:astro-ph/0405217
%\bibitem[Gould \& Loeb(1992)]{gouldloeb92} Gould, A. \& Loeb, A. 1992, \apj, 396, 104
%\bibitem[Gould et al.(2006)]{gould06} Gould, A., Udalski, A., An, D., et al.\ 2006, \apjl, 644, L37
%\bibitem[Gould et al.(2009)]{gould09} Gould, A., Udalski, A., Monard, B., et al.\ 2009, \apjl, 698, L147
\bibitem[Gould et al.(2014)]{gould14} Gould, A., Udalski, A., Shin, I.-G., et al.\ 2014, Science, 345, 46 
%\bibitem[Gould \& Yee(2013)]{gould_yee_terpar13} Gould, A., \& Yee, J.~C.\ 2013, \apj, 764, 107
%\bibitem[Green et al.(2012)]{WFIRST_rep} Green, J., Schechter, P., Baltay, C., et al.\ 2012, arXiv:1208.4012 
%\bibitem[Griest \& Hu(1992)]{griest92} Griest, K., \& Hu, W.\ 1992, \apj, 397, 362
%\bibitem[Griest \& Safizadeh(1998)]{griest98} Griest, K., \& Safizadeh, N.\ 1998, \apj, 500, 37
%\bibitem[Han et al.(2016)]{han_ob130723} Han, C., Bennett, D.~P., Udalski, A., \& Jung, Y.~K.\ 2016, \apj, in press (arXiv:1604.06533)
%\bibitem[Han \& Gould(1997)]{han97} Han , C., \& Gould, A.\ 1997, \apj, 480, 196
%\bibitem[Guillochon et al.(2011)]{guillochon11} Guillochon, J., Ramirez-Ruiz, E., \& Lin, D.\ 2011, \apj, 732, 74 
%\bibitem[Han et al.(2005)]{han_widepl2005} Han, C., Gaudi, B.~S., An, J.~H., \& Gould, A.\ 2005, \apj, 618, 962 
%\bibitem[Han \& Kang(2003)]{han_widepl2003} Han, C., \& Kang, Y.~W.\ 2003, \apj, 596, 1320
%\bibitem[Hartman et al.(2004)]{hartmanISIS} Hartman, J. D., Bakos, G., Stanek, K. Z., \& Noyes, R. W. 2004, AJ, 128, 1761
%\bibitem[Henderson et al.(2014)]{henderson14} Henderson, C.~B., Park, H., Sumi, T., et al.\ 2014, \apj, 794, 71
\bibitem[Henry et al.(1999)]{henry99} Henry, T.~J., Franz, O.~G., Wasserman, L.~H., et al.\ 1999, \apj, 512, 864 
\bibitem[Henry \& McCarthy(1993)]{henry93} Henry, T.~J., \& McCarthy, D.~W., Jr.\ 1993, \aj, 106, 773
\bibitem[Hirao et al.(2017)]{hirao17} Hirao, Y., Udalski, A., Sumi, T., et al.\ 2017, \aj, 154, 1 
%\bibitem[Hilton(2011)]{hilton11} Hilton, E.~J.\ 2011, PhD Thesis, University of Washington
%\bibitem[Hilton et al.(2010)]{hilton10} Hilton, E.~J., Hawley, S.~L., Kowalski, A.~F., \& Holtzman, J.\ 2010, arXiv:1012.0577 
%\bibitem[Holman \& Wiegert(1999)]{holman99} Holman, M.~J., \& Wiegert, P.~A.\ 1999, \aj, 117, 621 
%\bibitem[Heyrovsky(2003)]{heyrovsky03} Heyrovsk\'y, D.\ 2003, \apj, 594, 464
%\bibitem[Heyrovsky(2007)]{heyrovsky07} Heyrovsk\'y, D.\ 2007, \apj, 656, 483
\bibitem[Holtzman et al.(1998)]{holtzman98} Holtzman, J.~A., Watson, A.~M., Baum, W.~A., et al.\ 1998, \aj, 115, 1946 
%\bibitem[Howard et al.(2010)]{howard10}Howard, A.W. et al.\ 2010, Science, 330, 653
%\bibitem[Hubickyj et al.(2005)]{hubickyj05} Hubickyj, O., Bodenheimer, P., \& Lissauer, J.J.\ 2005, Icarus, 179, 415
\bibitem[Hwang et al.(2018)]{hwang18} Hwang, K.-H., Udalski, A., Shvartzvald, Y., et al.\ 2018, \aj, 155, 20
%\bibitem[Ida \& Lin(2004)]{idalin04} Ida, S.\ \& Lin, D.N.C.\ 2004, \apj, 604, 388
%\bibitem[Ida \& Lin(2005)]{ida05} Ida, S., \& Lin, D.N.C.\ 2005, \apj, 626, 1045
%\bibitem[Janczak et al.(2010)]{janczak10} Janczak, J., Fukui, A., Dong, S., et al.\ 2010, \apj, 711, 731
%\bibitem[Johnson et al.(2007)]{johnson07} Johnson, J.~A., Butler, R.~P., Marcy, G.~W., et al.\ 2007, \apj, 670, 833
%\bibitem[Johnson et al.(2010)]{johnson10} Johnson, J.~A., Aller, K.~M., Howard, A.~W., \& Crepp, J.~R.\ 2010, \pasp, 122, 905 
%\bibitem[Kaib et al.(2013)]{kaib13} Kaib, N.~A., Raymond, S.~N., \& Duncan, M.\ 2013, \nat, 493, 381 
%\bibitem[Kennedy \& Kenyon(2008)]{kennedy_snowline} Kennedy, G.~M., \& Kenyon, S.~J.\ 2008, \apj, 673, 502 
%\bibitem[Kennedy et al.(2006)]{kennedy-searth} Kennedy, G.M., Kenyon, S.J.,  \& Bromley, B.C.\ 2006, \apjl 650, L139
\bibitem[Kenyon \& Hartmann(1995)]{kenyon95} Kenyon, S.~J., \& Hartmann, L.\ 1995, \apjs, 101, 117 
\bibitem[Kervella et al.(2004)]{kervella_dwarf} Kervella, P., Th{\'e}venin, F., Di Folco, E., \& S{\'e}gransan, D.\ 2004, \aap, 426, 297
%\bibitem[Kervella et al.(2004)]{kervella04g} Kervella P., et al.\ 2004,  \aap, 428, 587
%\bibitem[Kervella \& Fouqu{\'e}(2008)]{kervella08} Kervella, P., \& Fouqu{\'e}, P.\ 2008, \aap, 491, 855
%\bibitem[Kim et al.(2013)]{kmtnet_pipe} Kim, D.~J., Lee, C.~U., Kim, S.~L., \& Park, B.~G.\ 2013, Pub.\ Korean Astron.\ Soc., 28, 1
%\bibitem[Kim et al.(2010)]{kmtnet10} Kim, S.-L., Park, B.-G., Lee, C.-U., et al.\ 2010, \procspie, 7733, 77733
\bibitem[Kim et al.(2016)]{kmtnet} Kim, S.-L., Lee, C.-U., Park, B.-G., et al.\ 2016, Journal of Korean Astronomical Society, 49, 37 
\bibitem[Koshimoto et al.(2017a)]{kosh17_ob120950} Koshimoto, N., Udalski, A., Beaulieu, J.~P., et al.\ 2017a, \aj, 153, 1 
\bibitem[Koshimoto et al.(2017b)]{kosh17_mb16227} Koshimoto, N., Shvartzvald, Y., Bennett, D.~P., et al.\ 2017b, \aj, 153, 1 
\bibitem[Koshimoto et al.(2014)]{koshimoto14} Koshimoto, N., Udalski, A., Sumi, T., et al.\ 2014, \apj, 788, 128
%\bibitem[Kowalski et al.(2009)]{kowalski09} Kowalski, A.~F., Hawley, S.~L., Hilton, E.~J., et al.\ 2009, \aj, 138, 633
%\bibitem[Kowalski et al.(2010)]{kowalski10} Kowalski, A.~F., Hawley, S.~L., Holtzman, J.~A., Wisniewski, J.~P., \& Hilton, E.~J.\ 2010, \apjl, 714, L98
%\bibitem[Koz{\l}owski et al.(2006)]{koz06} Koz{\l}owski, S., Wo{\'z}niak, P.~R., Mao, S., et al.\ 2006, \mnras, 370, 435
%\bibitem[Kubas et al.(2012)]{moa192_naco} Kubas, D., Beaulieu, J.~P., Bennett, D.~P., et al.\ 2012, \aap, 540, A78
%\bibitem[Kurucz(1993a)]{kurucz93a} Kurucz, R.L.\ 1993a, Kurucz CD-ROM 16, (SAO, Cambridge, MA, 1993).
%\bibitem[Kurucz(1993b)]{kurucz93b} Kurucz, R.L.\ 1993b, Kurucz CD-ROM 17,  (SAO, Cambridge, MA, 1993)
%\bibitem[Kurucz(1994)]{kurucz94} Kurucz, R.L.\ 1994, Kurucz CD-ROM 19,  (SAO, Cambridge, MA, 1993)
%\bibitem[Kurucz(1996)]{kurucz} Kurucz, R.L.\ 1996, ASP Conference Series, 108, 2
%\bibitem[Laughlin et al.(2004)]{laughlin04} Laughlin, G.  Bodenheimer, P.\ \& Adams, F.C.\ 2004, \apjl, 612, L73
%\bibitem[Lecar(2006)]{lecar_snowline} Lecar, M., Podolak, M., Sasselov, D., \& Chiang, E.\ 2006, \apj, 640, 1115
%\bibitem[Levison et~al.(1998)]{levison98} Levison, H.~F., Lissauer, J.~J., Duncan, M.~J.\ 1998, \aj, 116, 1998
%\bibitem[Lissauer(1993)]{lissauer_araa} Lissauer, J.J.\ 1993, Ann.\ Rev.\ Astron.\ Ast., 31, 129
%\bibitem[Malmberg et al.(2011)]{malmberg11} Malmberg, D., Davies, M.~B., \& Heggie, D.~C.\ 2011, \mnras, 411, 859
\bibitem[Lu et al.(2014)]{lu14} Lu, J.~R., Neichel, B., Anderson, J., et al.\ 2014, \procspie, 9148, 91480B 
\bibitem[Lu et al.(2016)]{lu16} Lu, J.~R., Sinukoff, E., Ofek, E.~O., Udalski, A., \& Kozlowski, S.\ 2016, \apj, 830, 41 
%\bibitem[Mao \& Paczy\'{n}ski(1991)]{mao91}Mao, S., \& Paczy\'{n}ski, B.\ 1991, \apj, 374, L37
%\bibitem[Mayor \& Queloz(2012)]{mayor12} Mayor, M., \& Queloz, D.\ 2012, \nar, 56, 19
%\bibitem[Minniti et al.(2010)]{minniti-vvv} Minniti, D., Lucas, P.~W., Emerson, J.~P., et al.\ 2010, \na, 15, 433 
%\bibitem[Musielak et al.(2005)]{musielak05} Musielak, Z.~E., Cuntz, M., Marshall, E.~A., \& Stuit, T.~D.\ 2005, \aap, 434, 355 
%\bibitem[Miyake et al.(2011)]{miyake11} Miyake, N., Sumi, T., Dong, S., et al.\ 2011, \apj, 728, 120
%\bibitem[Montet et al.(2013)]{montet13} Montet, B.T., Crepp, J.R., Johnson, J.A., Howard, A.W., \& Marcy, G.W.\ 2013, \apj, submitted (arXiv:1307.5849 )
%\bibitem[Movshovitz \& Podolak(2008)]{movshovitz08} Movshovitz, N., \& Podolak, M.\ 2008, Icarus, 194, 368
\bibitem[Mr{\'o}z et al.(2017a)]{mroz17_ob160596} Mr{\'o}z, P., Han, C., and, et al.\ 2017a, \aj, 153, 143 
\bibitem[Mr{\'o}z et al.(2017b)]{mroz17_2pl} Mr{\'o}z, P., Udalski, A., Bond, I.~A., et al.\ 2017b, \aj, 154, 205 
%\bibitem[Mullally et al.(2016)]{mullaly16} Mullally, F., Coughlin, J.~L., Thompson, S.~E., et al.\ 2016, arXiv:1602.03204 
%\bibitem[Muraki et al.(2011)]{muraki11} Muraki, Y., Han, C., Bennett, D.~P., et al.\ 2011, \apj, 741, 22
%\bibitem[Nagasawa \& Ida(2011)]{nagasawa11} Nagasawa, M., \& Ida, S.\ 2011, \apj, 742, 72 
\bibitem[Nataf et al.(2013)]{nataf13} Nataf, D.~M., Gould, A., Fouqu{\'e}, P., et al.\ 2013, \apj, 769, 88 
%\bibitem[Nishiyama et al.(2009)]{nish09} Nishiyama, S., Tamura, M., Hatano, H., et al.\ 2009, \apj, 696, 1407
%\bibitem[Park et al.(2012)]{kmtnet} Park, B.-G., Kim, S.-L., Lee, J.~W., et al.\ 2012, \procspie, 8444, 844447-1
%\bibitem[Pejcha \& Heyrovsky(2009)]{pei_hey} Pejcha, O., \& Heyrovsk\'y, D.\ 2009, \apj, 690, 1772
%\bibitem[Penny et al.(2013)]{penny13} Penny, M.~T., Kerins, E., Rattenbury, N., et al.\ 2013,  \mnras, submitted (arXiv:1206.5296)
\bibitem[Penny et al.(2018)]{penny18} Penny, M.~T., Gaudi, B.~S., Kerins, E., et al.\ 2018, in preparation
%\bibitem[Pietrukowicz et al.(2011)]{m22ml} Pietrukowicz, P., Minniti, D., Jetzer, P., Alonso-Garcia, J., \& Udalski, A.\ 2011, \apj, 744, L18
%\bibitem[Poindexter et al.(2005)]{poindexter05} Poindexter, S., et al.\ 2005, \apj, 633, 914
%\bibitem[Pollack et al.(1996)]{pollack96} Pollack, J.~B., Hubickyj, O., Bodenheimer, P., et al.\ 1996, \icarus, 124, 62
%\bibitem[Popowski et al.(2003)]{pop_extinct} Popowski, P., Cook, K.~H., \& Becker, A.~C.\ 2003, \aj, 126, 2910
%\bibitem[Quanz et al.(2012)]{quanz12} Quanz, S.~P., Lafreni{\`e}re, D., Meyer, M.~R., Reggiani, M.~M., \& Buenzli, E.\ 2012, \aap, 541, A133 
%\bibitem[Ranc et al.(2015)]{ranc15} Ranc, C., Cassan, A., Albrow, M.~D., et al.\ 2015, \aap, 580, A125
\bibitem[Ranc et al.(2018)]{ranc18} Ranc, C., et al.\ 2018, in preparation
%\bibitem[Rafikov(2011)]{rafikov11} Rafikov, R.\ 2011, \apj, 727, 86
%\bibitem[Rattenbury et al.(2007)]{rattenbury07} Rattenbury, N.~J., Mao, S., Debattista, V.~P., et al.\ 2007, \mnras, 378, 1165
\bibitem[Rattenbury et al.(2015)]{rattenbury15} Rattenbury, N.~J., Bennett, D.~P., Sumi, T., et al.\ 2015, \mnras, 454, 946 
\bibitem[Rattenbury et al.(2017)]{rattenbury17} Rattenbury, N.~J., Bennett, D.~P., Sumi, T., et al.\ 2017, \mnras, 466, 2710 
%\bibitem[Refsdal(1966)]{refsdal-par} Refsdal, S. 1966, \mnras, 134, 315
\bibitem[Rhie et al.(1999)]{rhie_98smc1} Rhie, S.~H., Becker, A.~C., Bennett, D.~P., et al.\ 1999, \apj, 522, 1037
%\bibitem[Rhie et al.(2000)]{rhie00} Rhie, S.~H., Bennett, D.~P., Becker, A.~C., et al.\ 2000, \apj, 533, 378
%\bibitem[Robin et al.(2003)]{robin03} Robin, A.~C., Reyl{\'e}, C., Derri{\`e}re, S., \& Picaud, S.\ 2003, \aap, 409, 523
\bibitem[Sako et al.(2008)]{sako_moacam3} Sako, T., Sekiguchi, T., Sasaki, M., et al.\ 2008, Experimental Astronomy, 22, 51
\bibitem[Schechter, Mateo, \& Saha(1993)]{dophot} Schechter, P.~L., Mateo, M., \& Saha, A.\ 1993, \pasp, 105, 1342
\bibitem[Shvartzvald et al.(2018)]{shvartzvald18} Shvartzvald, Y., Calchi Novati, S., Gaudi, B.~S., et al.\ 2018, \apjl, 857, L8 
\bibitem[Shvartzvald et al.(2016)]{shvartzvald16} Shvartzvald, Y., Maoz, D., Udalski, A., et al.\ 2016, \mnras, 457, 4089 
%\bibitem[Skowron(2011)]{skowron11} Skowron, J., et al.\ 2011, \apj, submitted, (arXiv:1101.3312)
%\bibitem[Skowron et al.(2013)]{skowron13} Skowron, J., et al.\ 2011, \apj, submitted
%\bibitem[Smith et al.(2002)]{smith02} Smith, M.C., Mao, S., \& Wo\'zniak, P.\ 2002, \mnras, 332, 962
%\bibitem[Smith et al.(2003)]{smith03} Smith, M.C., Mao, S., \& Paczy{\'n}ski, B.\ 2003, \mnras, 339, 925
\bibitem[Spergel et al.(2015)]{WFIRST_AFTA} Spergel, D., Gehrels, N., Baltay, C., et al.\ 2015, arXiv:1503.03757 
%\bibitem[Stetson(1994)]{allframe} Stetson, P.B.\ 1994, \pasp, 106, 250
\bibitem[Street et al.(2016)]{street16} Street, R.~A., Udalski, A., Calchi Novati, S., et al.\ 2016, \apj, 819, 93
%\bibitem[Stubbs et al.(2007)]{stubbs07} Stubbs, C.W., et al.\ 2007, \pasp, 119, 1163
%\bibitem[Sumi et al.(2010)]{sumi10}Sumi, T., Bennett, D.~P., Bond, I.~A. et al.\ 2010,  \apj, 710, 1641
%\bibitem[Sumi et al.(2011)]{sumi11}Sumi, T., Kamiya, K., Bennett, D.~P., et al.\ 2011,  \nat, 473, 349
\bibitem[Sumi et al.(2016)]{sumi16} Sumi, T., Udalski, A., Bennett, D.~P., et al.\ 2016, \apj, 825, 112 
%\bibitem[Sumi et al.(2004)]{sumi04} Sumi, T., Wu, X., Udalski, A., et al.\ 2004, \mnras, 348, 1439
\bibitem[Suzuki et al.(2016)]{suzuki16} Suzuki, D., Bennett, D.~P., Sumi, T., et al.\ 2016, \apj, 833, 145
\bibitem[Suzuki et al.(2014)]{suzuki14} Suzuki, D., Udalski, A., Sumi, T., et al.\ 2014, \apj, 780, 123 
\bibitem[Suzuki et al.(2014e)]{suzuki14e} Suzuki, D., Udalski, A., Sumi, T., et al.\ 2014e, \apj, 788, 97 
\bibitem[Szyma{\'n}ski et al.(2011)]{ogle3-phot} Szyma{\'n}ski, M.~K., Udalski, A., Soszy{\'n}ski, I., et al.\ 2011, \actaa, 61, 83 
\bibitem[Tang et al.(2014)]{tang14_PARSEC} Tang, J., Bressan, A., Rosenfield, P., et al.\ 2014, \mnras, 445, 4287 
%\bibitem[Thommes et al.(2008)]{thommes08} Thommes, E.W., Matsumura, S., \& Rasio F.A.\ 2008, Science, 321, 814
\bibitem[Tomaney \& Crotts (1996)]{tom96}Tomaney, A.B. \& Crotts, A.P.S.\ 1996, \aj 112, 2872
%\bibitem[Traub(2011)]{traub11} Traub, W.\ 2011, ApJ, submitted.
%\bibitem[Twicken et al.(2016)]{kepler_q17} Twicken, J.~D., Jenkins, J.~M., Seader, S.~E., et al.\ 2016, ApJ, submitted (arXiv:1604.06140)
%\bibitem[Udalski(2003a)]{udalski-ext} Udalski, A.\ 2003a, \apj, 590, 284
\bibitem[Udalski(2003)]{ogle-pipeline} Udalski, A.\ 2003, \actaa, 53, 291
%\bibitem[Udalski et al.(2005)]{udalski05} Udalski, A., Jaroszy{\'n}ski, M., Paczy{\'n}ski, B., et al.\ 2005, \apjl, 628, L109
%\bibitem[Udalski et al.(2015b)]{udalski_ob130723} Udalski, A., Jung, Y.~K., Han, C., et al.\ 2015b, \apj, 812, 47 
%\bibitem[Udalski et al.(1994)]{ogle-ews} Udalski, A., Szyma\'{n}ski, M., Ka{\l}u\.{z}ny, J., Kubiak, M., Mateo, M., Krzmi\'{n}ski, W., \& \pac, B.\ 1994, Acta Astron., 44, 227
%\bibitem[Udalski et al.(2002)]{uda02} Udalski, A., Szymanski, M., Kubiak, M., et al.\ 2002, \actaa, 52, 217
%\bibitem[Udalski et al.(2008)]{udalski08} Udalski, A., et al.\ 2008, Acta Astron., 58, 69
\bibitem[Udalski et al.(2015a)]{ogle4} Udalski, A., Szyma{\'n}ski, M.~K., \& Szyma{\'n}ski, G.\ 2015a, \actaa, 65, 1 
%\bibitem[Udalski et al.(2018)]{udalski18} Udalski, A., Ryu, Y.-H., Sajadian, S., et al.\ 2018, \actaa, 68, 1 
%\bibitem[Veras et al.(2011)]{veras11}  Veras, D., Wyatt, M.~C., Mustill, A.~J., Bonsor, A., \& Eldridge, J.~J.\ 2011, \mnras, 417, 2104
%\bibitem[Veras \& Raymond(2012)]{veras12} Veras, D., \& Raymond, S.~N.\ 2012, \mnras, 421, L117
%\bibitem[Veras \& Tout(2012)]{veras_tout12} Veras, D., \& Tout, C.~A.\ 2012, \mnras, 422, 1648 
%\bibitem[Voyatzis et al.(2013)]{voyatzis13} Voyatzis, G., Hadjidemetriou, J.~D., Veras, D., \& Varvoglis, H.\ 2013, \mnras, 430, 3383 
%\bibitem[Wambsganss(2011)]{wamb11} Wambsganss, J.\ 2011, \nat, 473, 289
%\bibitem[Ward(1997)]{ward97} Ward, W.R.\ 1997, Icarus, 126, 261
%\bibitem[Witt(1995)]{witt95} Witt, H.J.\ 1995, \apj, 449, 42
%\bibitem[Wright \& Gaudi(2013)]{wright_gaudi_book2013} Wright, J.~T., \& Gaudi, B.~S.\ 2013, Planets, Stars and Stellar Systems.~Volume 3: Solar and Stellar Planetary Systems, 489 
%\bibitem[Yee(2013)]{yee_WFIRST_par} Yee, J.~C.\ 2013, \apjl, 770, L31 
\bibitem[Yee et al.(2012)]{yee12} Yee, J.~C., Shvartzvald, Y., Gal-Yam, A., et al.\ 2012, \apj, 755, 102
%\bibitem[Yee et al.(2009)]{yee09} Yee, J.~C., Udalski, A., Sumi, T., et al.\ 2009, \apj, 703, 2082 
\bibitem[Yoo et al.(2004)]{yoo_rad} Yoo, J.~et al.\ 2004,  \apj, 603, 139
\end{thebibliography}
\end{document}